\documentclass[prodmode,acmjetc]{acmsmall} % Aptara syntax

\usepackage[ruled]{algorithm2e}
\usepackage{amsmath}
\usepackage{enumerate}
\renewcommand{\algorithmcfname}{ALGORITHM}
\SetAlFnt{\small}
\SetAlCapFnt{\small}
\SetAlCapNameFnt{\small}
\SetAlCapHSkip{0pt}
\IncMargin{-\parindent}

\acmVolume{}
\acmNumber{}
\acmArticle{}
\acmYear{}
\acmMonth{3}

\doi{0000001.0000001}

\issn{1234-56789}

\begin{document}

% Page heads
\markboth{C.-H. Chien et al.}{Fault-tolerant Operations for Universal Blind Quantum Computation}

% Title portion
\title{Fault-tolerant Operations for Universal Blind Quantum Computation}
\author{CHIA-HUNG CHIEN
\affil{National Taiwan University}
RODNEY VAN METER
\affil{Keio University}
SY-YEN KUO
\affil{National Taiwan University}}
% NOTE! Affiliations placed here should be for the institution where the
%       BULK of the research was done. If the author has gone to a new
%       institution, before publication, the (above) affiliation should NOT be changed.
%       The authors 'current' address may be given in the "Author's addresses:" block (below).
%       So for example, Mr. Abdelzaher, the bulk of the research was done at UIUC, and he is
%       currently affiliated with NASA.

\begin{abstract}
Blind quantum computation is an appealing use of quantum information technology because it can conceal both the client's data and the algorithm itself from the server.  However, problems need to be solved in the practical use of blind quantum computation and fault-tolerance is a major challenge. On an example circuit, the computational cost measured in T gates executed by the client is 97 times more than performing the original computation directly, without using the server, even before applying error correction. (The client still benefits due to drastically reduced memory requirements.) Broadbent et al. proposed running error correction over blind computation, but our first protocol applies one layer of Steane's [[7,1,3]] code underneath instead. This protocol has better fault tolerance, but still results in a substantial overhead.  We propose another protocol to reduce the client's computational load by transferring the qubit preparation to the server. For each logical qubit used in the computation, the client is only required to receive eight logical qubits via teleportation then buffer two logical qubits before returning one.  This protocol also protects the client's fault-tolerant preparation of logical qubits from a side-channel attack.
\end{abstract}




\terms{Design, Reliability}


\acmformat{Chia-Hung Chien, Rodney Van Meter, Sy-Yen Kuo, 2013. Fault-tolerant operations for universal blind quantum computation.}
% At a minimum you need to supply the author names, year and a title.
% IMPORTANT:
% Full first names whenever they are known, surname last, followed by a period.
% In the case of two authors, 'and' is placed between them.
% In the case of three or more authors, the serial comma is used, that is, all author names
% except the last one but including the penultimate author's name are followed by a comma,
% and then 'and' is placed before the final author's name.
% If only first and middle initials are known, then each initial
% is followed by a period and they are separated by a space.
% The remaining information (journal title, volume, article number, date, etc.) is 'auto-generated'.

\begin{bottomstuff}
This work is supported by National Science Council, Taiwan under Grant NSC 99-2221-E-002-106-MY3, and ``111" Project under Grant B08002. RV acknowledges support from the Japan Society for the Promotion of Science (JSPS) through its Funding Program for World-Leading Innovation R\&D on Science and Technology (FIRST Program).

Author's addresses: C.-H. Chien and S.-Y. Kuo, Department of Electrical Engineering, National Taiwan University, Taiwan; R. Van Meter, Faculty of Environment and Information Studies, Keio University, Japan.
\end{bottomstuff}

\maketitle

\section{Introduction}
%Motivation
%Blind quantum computation is good but require fault-tolerant
%computation error device imperfection noise on channel
Quantum computation shows great potential for solving some important problems faster than classical computation \cite{Mosca08,BD10,Kassal11,Cody12}. However, a practical quantum computer needs to be large enough to handle sufficient numbers of delicate qubits, perhaps extending into the high millions or low billions of physical qubits \cite{rdv10,Jones12}. Large-scale quantum ``mainframes" will be valuable resources, and time-sharing of machines will be economically attractive. Time-sharing quantum cloud services will allow owners of smaller quantum computers to perform large quantum computations. Sometimes the input and output data are private and even the choice of quantum computing algorithm may be sensitive information, so this information has to be kept secret even from the server. Universal blind quantum computation is a powerful technique that can let the client do a quantum computation on a server without revealing any information about the computation except an upper bound on the size \cite{BFK09}, akin to classical homomorphic encryption but with a much more tractable performance penalty relative to an unencrypted computation.

Universal blind quantum computation makes use of a special feature of measurement-based quantum computation \cite{RB01,RBB03} which allows a quantum computation to be divided into quantum operations and classical computations. Measurement-based quantum computation uses entangled qubits as resources and the computation is performed only by a sequence of measurements. The type of measurements determines the quantum operation on the output qubits. In blind quantum computation, the client prepares qubits and decides the type of measurements while the server does most of the quantum operations. The inputs are encrypted in the qubits prepared by the client, and the measurements are also encrypted so that they are independent of the real computation. Thus, from the server's point of view, all the received quantum and classical information appears random. Blind quantum computation with four qubits was experimentally demonstrated in 2012 \cite{Barz12}.

%fault-tolerant quantum computation requires a lot of resources
%evaluate the resources is an important question of implementing practical system
A practical blind quantum computing system requires the ability to operate in the presence of errors. Errors may occur during preparation of qubits, computation, and transmission the qubits because of imperfect computing devices or a noisy transmission channel. Broadbent {\em et al.} suggested doing a fault-tolerant quantum computation as a target computation in their scheme. The error correction itself, including syndrome measurement and corrections, would be performed as a blind computation, necessitating enormous amounts of low-latency, near-real-time classical communication. It also require a teleportation rate sufficient to support real-time quantum error correction, rendering the scheme effectively impossible. However, in this scheme, quantum error correction is only applied to the server. If the client and the server are connected by a long and noisy channel, or more likely a quantum repeater network \cite{Dur99,GR07,Kimble08,Rodney12}, it will be hard to correct the errors after the server receives the qubits. Blind quantum computation will be limited to using in a short range. Thus, we suggest applying quantum error correction to the client, too, and applying it underneath the blinding protocol.

% advantage sequence
Based on Broadbent {\em et al.}'s approach, we propose our first fault-tolerant blind quantum computation protocol, which uses a quantum error correction code and fault-tolerant quantum circuits. The client has to do fault-tolerant quantum computation to prepare logical qubits encoded with a quantum error correction code, and the server has to conduct fault-tolerant entanglements and measurements on the encoded logical qubits. Compared to Broadbent {\em et al.}'s fault-tolerant protocol, our first protocol provides a better fault-tolerant capability because errors that occur during qubit preparation and teleportation can be corrected. In addition, their fault-tolerant protocol applies quantum error correction \emph{on top of} the basic blind quantum computation, while our fault-tolerant protocol applies quantum error correction \emph{underneath} the basic blind quantum computation. In comparison to Broadbent {\em et al.}'s basic blind quantum computation protocol, the number of qubits used in our first fault-tolerant protocol is increased by a constant factor, while the number of qubits used in their fault-tolerant protocol has a linear growth. A major drawback of our first fault-tolerant protocol is that the client is required to do a lot of quantum computations for the fault-tolerant preparation of encoded logical qubits. Since the client is assumed to have limited quantum computing power, our first protocol needs to be improved. In our first fault-tolerant protocol, the client may also leak some information which can be exploited by a side-channel attack.

%Alice drive the computation
To reduce the client's large quantum computing overhead, we propose our second fault-tolerant blind quantum computation protocol. The error-correcting qubits will be prepared fault-tolerantly by the server. The client is required only to buffer eight logical qubits and send one chosen at random every time when generating an input logical qubit. Our second fault-tolerant protocol avoids the additional quantum computing requirement for the client and still preserves two advantages of our first protocol. It provides more fault-tolerance and uses fewer qubits. Furthermore, the side-channel attack which works against our first protocol does not work against our second protocol. The idea of letting the server prepare error-correcting qubits can be applied to other protocol to make them more fault-tolerant.

%Since the client has limited quantum power, evaluating the overhead of the client in a fault-tolerant blind quantum computation is an important issue. We consider a scenario in which the client encodes every qubit with one layer of the Steane's [[7,1,3]] error correction code [Ref] and do a 10-qubit quantum carry-lookahead adder [Ref].
% It turns out the quantum computing overhead of the client is above 40 times more than the client doing by himself, although the client doesn't need as large quantum computer as doing by himself.
%Evaluation of resource required in a practical computation is important for implementing the system. We use 
%gives the reader the information that he or she needs to apply the results themselves to things they are considering

This paper is the first work analysing the quantum computation overhead of the client. In section 2, we will introduce related work. In section 3, we will describe the first protocol of fault-tolerant blind quantum computation using fault-tolerant circuit. In section 4, we will present the second protocol to reduce the client's computation effort. In section 5, we will analyse and compare the computation overhead between different protocols. Section 6 will be the conclusion.

\section{Related Work}
Protecting the client's private data from the computing server during a computation is a long-standing problem, but if achievable it is valuable for cloud computing. In 2009, this problem was solved both in classical computing and quantum computing. Gentry proposed a fully homomorphic encryption which makes it possible to perform mathematical operations directly on encrypted data and get the result in an encrypted form which can be decrypted only by someone holding the original encryption key \cite{Childs05,homenc,Gentry09}. Broadbent {\em et al.} proposed a universal blind quantum computation which allows the client to utilize the quantum computing power on the server without revealing the computation, including input, output and even the algorithm. The security of classical homomorphic encryption is based on difficult mathematical problems, while the security of the blind quantum computation is based on the physical properties of quantum systems, assuming the client has a good source of classical random bits. Homomorphic encryption has enormous computing overhead and the server learns what computation is being performed, even though it learns nothing about the data. Blind quantum computation requires sophisticated quantum computers. Though blind quantum computation is proved to be theoretically perfectly secure, the security of a practical implementation is an open question.

Childs first proposed the idea of blind quantum computation and proved that the client can conceal the quantum input and output from the server \cite{Childs05}. But the protocol requires the client to have a fault-tolerant quantum memory and to do quantum operations. Arrighi and Salvail proposed another approach which lets the server compute multiple quantum inputs, most of which are decoys \cite{AS06}. This approach can't do universal computation and the real quantum inputs are not encrypted. Broadbent {\em et al.} proposed the first universal blind quantum computation using the measurement-based quantum computation model \cite{RBB03,BFK09}. The client is only required to prepare a large set of qubits one at a time, randomly chosen from a finite set, and to send the qubits to the server. Dunjko {\em et al.} pointed out the difficulties of preparing and sending qubits over a long distance \cite{Dunjko12}. They proposed a remote blind qubit state preparation protocol that requires Alice to prepare only weak coherent pulses, but the blindness will not be perfect. Morimae and Fujii show a fault-tolerant blind quantum computation using topologically protected measurement-based quantum computation \cite{MF12}. Their approach uses a more fault-tolerant computation model on the server, but the input qubits are still not protected during the transmission. Other different blind quantum computation schemes are also proposed recently \cite{CCCK12,SKM13}. Compared to the previous protocol, our protocol applies quantum error correction on the client so that the input qubits can be protected during the transmission, while the client still has a low resource requirement. Our work covers three areas in quantum computation, including measurement-based quantum computation, universal blind quantum computation, and quantum error correction. We will introduce the basic concept of quantum computation first. Then we will describe each area in the following subsections.

\subsection{Basic Quantum Computation}
A qubit is a basic unit of quantum information, which is stored on a two-state quantum-mechanical system, e.g. the polarization of a single photon or the spin of a single electron \cite{NC10,LJLMNO10}. A qubit is a superposition of two basis states, and it is generally represented using the ket notation as $|\psi\rangle = \alpha|0\rangle + \beta|1\rangle$, where $\alpha$ and $\beta$ are probability amplitudes, and $\alpha, \beta \in \mathbb{C}$. $|0\rangle$ and $|1\rangle$ are ket notations for the basis vectors $\begin{pmatrix} 1 \\ 0 \end{pmatrix}$ and $\begin{pmatrix} 0 \\ 1 \end{pmatrix}$. When we measure this qubit in the computational (standard) basis $M_{Z}=\{|0\rangle,|1\rangle\}$, the outcome is $|0\rangle$ with probability $|\alpha|^2$ and $|1\rangle$ with probability $|\beta|^2$. More precisely, $(\alpha, \beta)$ can be represented as $(\cos(\theta/2), e^{i\phi}\sin(\theta/2))$, where $0 \leq \theta,\phi < 2\pi$, and $\phi$ is called the phase.

Two or more qubits can be entangled, and they will be in a superposition of a single quantum state, which means we can only know the state of the whole system but not any one of the qubits individually. For example, a common pair of qubits called a Bell pair or EPR pair is represented as $|\Phi^{+}\rangle = \frac{1}{\sqrt{2}}(|0\rangle_{A}\otimes|0\rangle_{B}+|1\rangle_{A}\otimes|1\rangle_{B})$, where the subscript is the index of qubits. If we measure the first qubit of the entangled states $|\Phi^{+}\rangle$ and the measurement result is $|1\rangle$, the second qubit will become $|1\rangle$. It seems that the two qubits in the entangled state have a connection. Although the state of each qubit is random, they are not independent. Their states are correlated in a fashion not explainable by sheer classical probability.

\begin{figure}
\centerline{\scalebox{.4}{\includegraphics{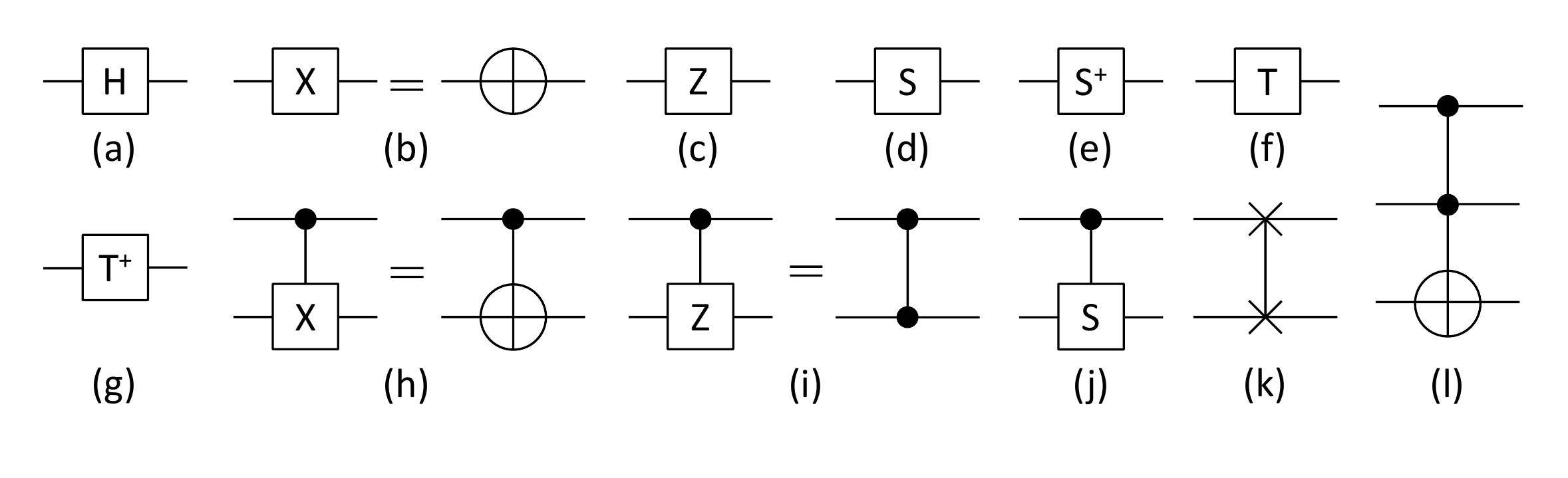}}}
\caption{Diagram of quantum gates. Each line (wire) represents a qubit. Input from the left and output from the right. In two-qubit and three-qubit gates, the wire with black dot represents a control qubit and the other represents a target qubit. (a) Hadamard gate (b) Pauli-X gate (c) Pauli-Z gate (d) S (phase) gate (e) inverse S gate (f) T ($\pi/8$) gate (g) inverse  T gate (h) CNOT gate (i) CZ gate (j) CPhase gate (k) SWAP gate (l) Toffoli gate}
\label{fig:gates}
\end{figure}

In general, a quantum computation is described using the quantum circuit model in which a computation is a sequence of quantum gates. Quantum gates are unitary operators generally represented by matrices. We can use the vector form to represent the qubit as $|\psi\rangle = \alpha\begin{pmatrix} 1 \\ 0 \end{pmatrix} + \beta\begin{pmatrix} 0 \\ 1 \end{pmatrix}$ when we want to calculate the output of the quantum gates. The following equation shows the matrix form of some quantum gates.
%Equation (\ref{eqn:01})
\begin{equation}
\label{eqn:01}
\begin{aligned}
&I=\left[
\begin{array}{cc}
1 & 0 \\
0 & 1
\end{array}\right],
H=\frac{1}{\sqrt{2}}\left[
\begin{array}{cc}
1 & 1 \\
1 & -1
\end{array}\right],
X=\left[
\begin{array}{cc}
0 & 1 \\
1 & 0
\end{array}\right],
Z=\left[
\begin{array}{cc}
1 & 0 \\
0 & -1
\end{array}\right],
R_{z}(\theta)=\left[
\begin{array}{cc}
1 & 0 \\
0 & e^{i\theta}
\end{array}\right], \\
&CNOT=\left[
\begin{array}{cccc}
1 & 0 & 0 & 0 \\
0 & 1 & 0 & 0 \\
0 & 0 & 0 & 1 \\
0 & 0 & 1 & 0
\end{array}\right],
CZ=\left[
\begin{array}{cccc}
1 & 0 & 0 & 0 \\
0 & 1 & 0 & 0 \\
0 & 0 & 1 & 0 \\
0 & 0 & 0 & -1
\end{array}\right],
CPhase=\left[
\begin{array}{cccc}
1 & 0 & 0 & 0 \\
0 & 1 & 0 & 0 \\
0 & 0 & 1 & 0 \\
0 & 0 & 0 & i
\end{array}\right],
\end{aligned}
\end{equation}
I is the identity gate. H is the Hadamard gate. X is the Pauli-X or NOT gate. Z is the Pauli-Z gate. $R_{z}(\theta)$ is the phase shift gate, where $0 \leq \theta < 2\pi$. $R_{z}(\pi)$ is the Pauli-Z gate, $R_{z}(\frac{\pi}{2})$ is called the S or phase gate, and $R_{z}(\frac{\pi}{4})$ is called the T gate. $R_{z}(\frac{3\pi}{2})$ is the inverse S gate, and $R_{z}(\frac{7\pi}{4})$ is the inverse T gate. CNOT is the controlled-NOT gate. CZ is the controlled-Z gate. CPhase is the controlled-phase gate. Fig.~\ref{fig:gates} shows the diagram of some gates in the quantum circuit model. Each line, called a wire, represents a qubit. The SWAP gate swaps two input qubits. For the CNOT, CZ, and CPhase gates, the upper qubit is the control qubit and the lower qubit is the target qubit. If the control bit is 1, CNOT will apply an X gate on the target qubit, CZ will apply a Z gate on the target qubit, and CPhase will apply an S gate on the target qubit. The Toffoli gate has two control qubits and one target qubit. Only when both control bits are 1, the Toffoli gate will apply an X gate on the target.

A qubit has an infinite number of possible states. When we measure a qubit with an operator $M$, which is a $2 \times 2$ Hermitian (self-adjoint) matrix, the qubit will be projected to one of its two eigenstates (eigenvectors) corresponding to the eigenvalue, which is the measurement result. The eigenstates of the operator form an orthonormal basis, thus, we can also use this basis to distinguish the measurement. In quantum computation, we generally measure a qubit in the computational basis $\{|0\rangle, |1\rangle\}$, which is the same as measuring with operator $Z$. Any qubit can be represented as a superposition of the eigenstates of the measurement operator. For example,
\begin{equation}
\label{eqn:02}
|\psi\rangle = \alpha|0\rangle + \beta|1\rangle = \frac{\alpha + \beta}{\sqrt{2}}(|0\rangle + |1\rangle) + \frac{\alpha - \beta}{\sqrt{2}}(|0\rangle - |1\rangle).
\end{equation}
When we measure a qubit $|\psi\rangle$ in the basis $\{|0\rangle \pm |1\rangle\}$, which is measuring with operator $X$, the qubit will be $\frac{1}{\sqrt{2}}(|0\rangle + |1\rangle)$ with probability $\frac{(\alpha + \beta)^2}{2}$ and $\frac{1}{\sqrt{2}}(|0\rangle + |1\rangle)$ with probability $\frac{(\alpha - \beta)^2}{2}$.

%A qubit can be represented as a superposition of the eigenvectors of the measurement operator. $|\psi\rangle = \Sigma_{i} c_{i}|\phi\rangle_{i}$

%MBQC
%a cluster state is a highly entangled state of qubits which can serve as the resource for universal quantum computation
%Measurement-based or one-way quantum computation is a method of quantum computing that first prepares an entangled resource state, usually a cluster state or graph state, then performs single qubit measurements on it. It is "one-way" because the resource state is destroyed by the measurements.
\subsection{Measurement-based Quantum Computation}
%Raussendorf's MBQC
Measurement-based quantum computation (MBQC) or one-way quantum computation is a quantum computation model proposed by Raussendorf and Briegel \cite{RB01,RBB03}. MBQC uses an entangled state of many qubits, called a cluster state, as the resource for universal quantum computation. The computation is a sequence of one-qubit measurements on the qubits of this cluster state. Every measurement will destroy the entanglement connection between the measured qubits and the others, and the remaining qubits become a smaller cluster state modified by an operation. The computation is driven by measurements in designated bases. Depending on the measurement basis and the measurement result, we learn something about the state of the remaining entangled qubits. Managed carefully, ultimately the resulting state represents the desired result, a function of our input qubits \cite{DK06}.

\begin{figure}
\centerline{\scalebox{.4}{\includegraphics{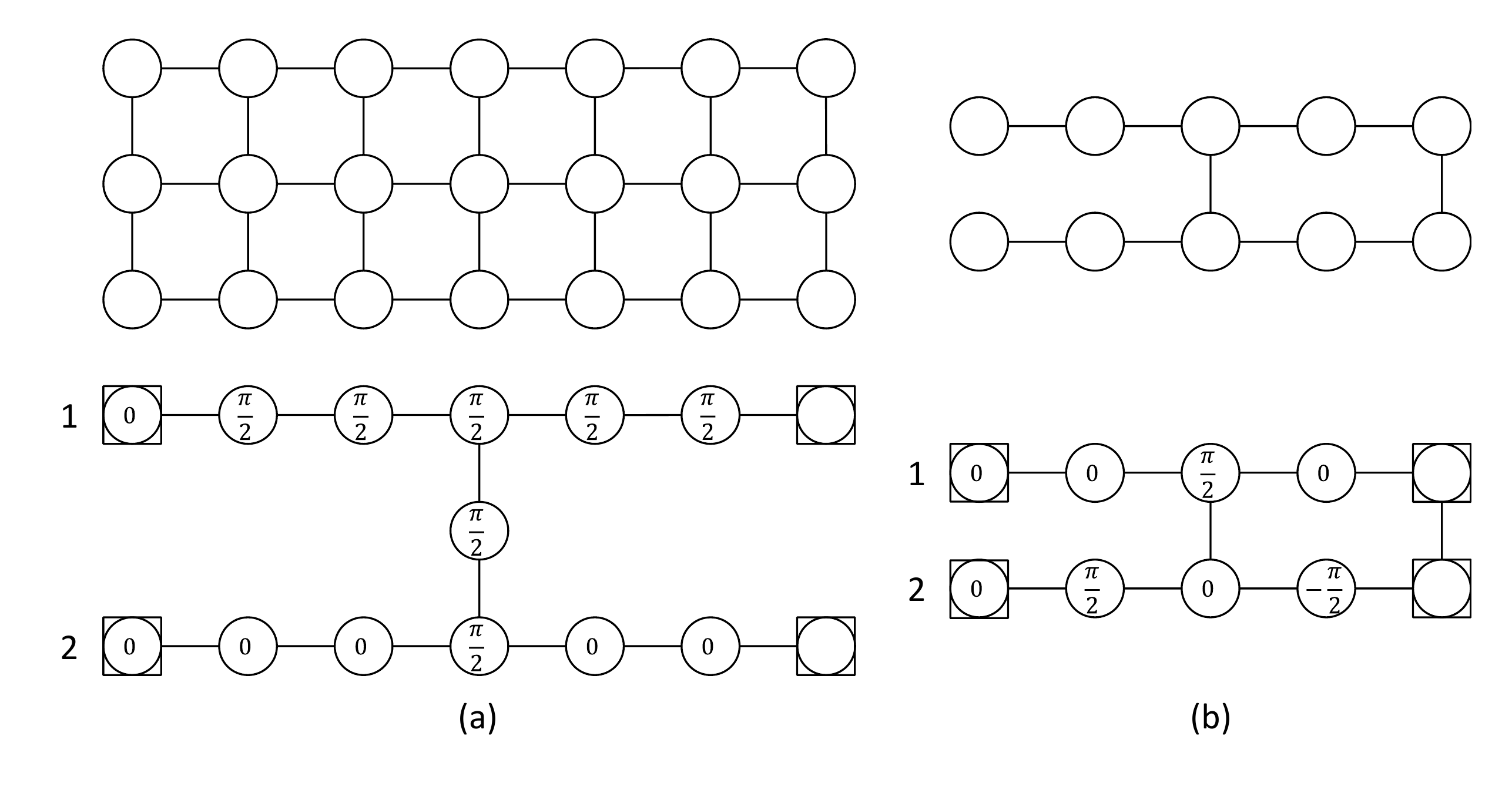}}}
\caption{Diagram of two measurement-based quantum computation using different entangled qubits. Circles represent qubits, and lines represent entangling operators. After the generic entangling operations, each qubit is measured in a carefully chosen basis. The value written in each circle is the angle of the measurement basis used for measuring the qubit. (a) Measurement-based quantum computation of a controlled-NOT gate on a 2-D cluster state. The removed qubits in the lower part is measured by the computational basis. (b) Measurement-based quantum computation of a controlled-NOT gate on a brick state.}
\label{fig:MBQCcomp}
\end{figure}

To prepare a cluster state, we initially prepare all the qubits in $|+\rangle=\frac{1}{\sqrt{2}}(|0\rangle+|1\rangle)$ state. Then we apply CZ to each nearest neighbour pair. A cluster state can have different structure, e.g. linear, 2-D, or 3-D. To do a computation on the cluster state, we imprint its quantum circuit layout on the substrate cluster state by the measurements. The redundant qubits are removed from the cluster state by being measured in the computational basis. Each remaining qubit is measured in a particular kind of basis $M(\theta)=\{|0\rangle\pm e^{i\theta}|1\rangle\}$, where $0 \leq \theta < 2\pi$, based on the quantum circuit. $\theta$ is called the angle of the measurement basis. The upper part of Fig.~\ref{fig:MBQCcomp}(a) shows a cluster state, which is a 2-dimensional array of cluster state. The lower part of Fig.~\ref{fig:MBQCcomp}(a) shows the angle $\theta$ of the measurement basis $M_{\theta}$ of executing a CNOT gate on two qubits using the 2-D cluster. A horizontal line of qubits is analogous to a wire in the quantum circuit diagram, see Fig.~\ref{fig:gates}. The two qubits in the left squares are the inputs, and two qubits in the right squares are the outputs; the upper qubits are the control qubits and the lower qubits are the target qubits. The output qubits have to be corrected by some Pauli gates based on the previous measurement results. The cluster state is generic and can be prepared before the computation, but the redundant qubits can only be removed after the target computation is chosen.

\subsection{Universal Blind Quantum Computation}
\begin{figure}
\centerline{\scalebox{.4}{\includegraphics{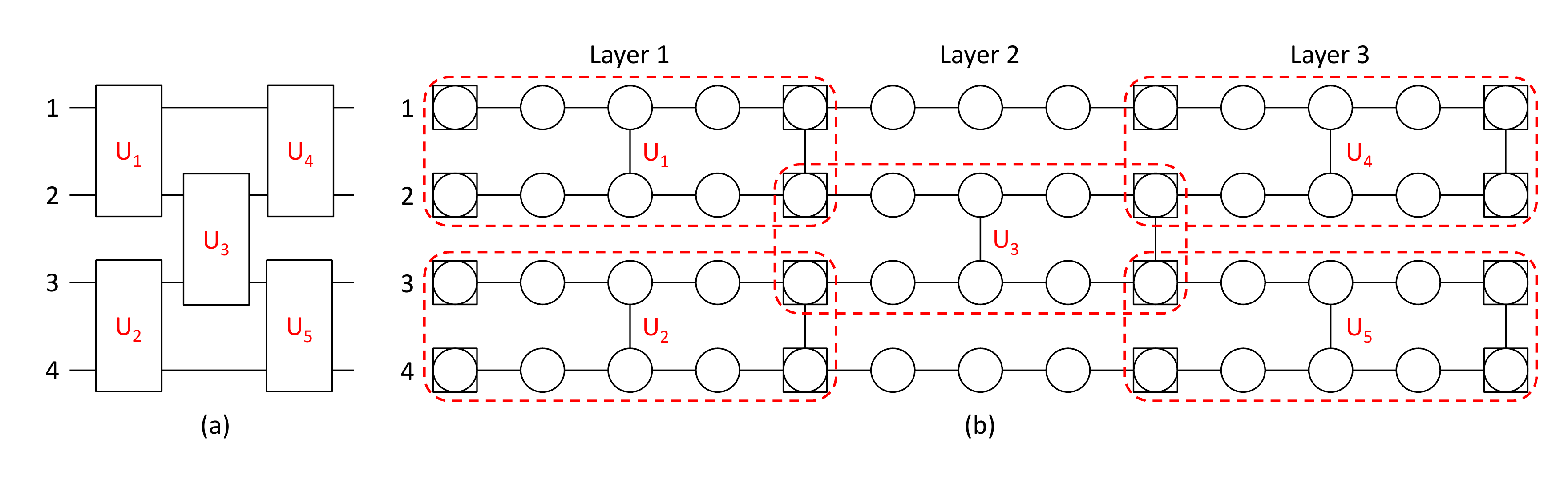}}}
\caption{(a) The circuit model of a quantum computation of four qubits. Each block represents a quantum gate. (b) The brickwork state for the quantum computation. The Five bricks are even-odd interleavingly tiled to a brickwork state with three layers. Each brick can be used to implement a quantum gate. In each brick, the squares on the left indicate inputs, and the squares on the right indicate outputs.}
\label{fig:BFK}
\end{figure}

%simple here
%measurement angle $\phi_{x,y}$
%The implicit required corrections are implemented according to the flow condition [DK06] which guarantees determinism, and allows measurements to be performed layer-by-layer. We now show how we can tile the patterns as given in Fig4 through 8. (The underlying graph states are the same) to implement any circuit using U as a universal set of gates. In fig9, we show how a 4-qubit circuit with three gates, U1, U2, and U3 (each acting on a maximum of two adjacent qubits) can be implemented on the brickwork state.
%(FT)the preparation of the input state is included in the computational part.
Universal blind quantum computation (UBQC) is a special application of measurement-based quantum computation. A client called Alice wants to execute a quantum computation on a server called Bob because she doesn't have enough quantum computing power. If Alice wants to conceal her computation from everyone else, including Bob, she can use universal blind quantum computation. Most of the blind quantum computation protocols are similar to the protocol of Broadbent, Fitzsimons, and Kashefi, which we call BFK. In order to conceal the computation, blind quantum computation does not use the typical cluster state because removing the redundant qubits reveals information about the choice of quantum gates, which is shown in Fig.~\ref{fig:MBQCcomp}(a). In standard MBQC, the computing qubits are measured in the $M(\theta)$ basis, while the redundant qubits are measured in the computational basis. BFK instead uses a specially designed entangled state called the brickwork state as shown in Fig.~\ref{fig:BFK}(b), where all qubits are measured in the $M(\theta)$ basis. The comparison of entangled states used in MBQC and UBQC is shown in Fig.~\ref{fig:MBQCcomp}, where the upper part shows the entangled states required to implement a CNOT gate, and the lower part shows the measurement basis to perform a CNOT gate. A brickwork state is a uniform structure, which only reveals the upper bound of the size of the computation. It is a tiling of 10-qubit entangled states, which we call bricks. Each brick can implement a quantum gate, including Hadamard gate, Pauli-X gate, Pauli-Z gate, S gate, T gate, and CNOT gate. This set is sufficient for universal quantum computation \cite{NC10}. We should note that bricks are interleaved, so quantum gates can only act on specific two adjacent lines of qubits in each layer of the brickwork state. Thus, swap gates are required for implementing quantum gates which are operated on two non-adjacent qubits in the blind quantum computation.

In measurement-based quantum computation, the output qubits have to be corrected by some Pauli gates based on the previous measurement results. The correction can be performed on the next measurement by the modification of measurement basis. For a brickwork state of the size $n \times m$, let each qubit $|\psi_{x,y}\rangle$ be indexed by a column $x \in {1,...n}$ and a row $y \in {1,...,m}$. Every measurement depends on the previous measurement results, and these results can be assigned to two dependency sets $X_{x,y}$ and $Z_{x,y}$ by the flow construction \cite{DK06}. The measurement basis with correction is $\phi'_{x,y} = (-1)^{s_{x,y}^X}\phi_{x,y} + s_{x,y}^{Z}\pi$, where $\phi_{x,y}$ is the original measurement basis for a quantum gate when the qubit is prepared as $|+\rangle$. 
If the qubits in the brickwork state are all prepared in the $|+\rangle$ state, then the measurement bases will reveal information about the quantum gate. Thus, each qubit is prepared in a randomly-chosen state from the set $\{\frac{1}{\sqrt{2}}(|0\rangle + e^{i\theta_{x,y}}|1\rangle) | \theta_{x,y} = 0,\pi/4,...,7\pi/4\}$, where $\theta_{x,y}$ is called the angle of the qubit. In measurement-base quantum computation, measuring a qubit $|0\rangle + e^{i\theta}|1\rangle$ in $M(\theta)$ has the same effect as measuring a qubit $|0\rangle + |1\rangle$ in $M(0)$. Therefore, the measurement basis of each qubit in the brickwork state becomes $M(\phi'_{x,y} + \theta_{x,y})$. We can see that the measurement basis will be random if $\theta_{x,y}$ is random. A random number is also added to the angle of the measurement basis to conceal the quantum information of the qubit, which will also flip the measurement results.
The basic steps of BFK's protocol are as follows.
\begin{enumerate}
  \item The preparation stage
  \begin{enumerate}
    \item Alice prepares each qubit $|\psi_{x,y}\rangle$ randomly chosen from the set $\{\frac{1}{\sqrt{2}}(|0\rangle + e^{i\theta_{x,y}}|1\rangle) | \theta_{x,y} = 0,\pi/4,...,7\pi/4\}$, and sends them to Bob via the quantum channel.
    \item Bob creates the entangled brickwork state shown in Fig.~\ref{fig:BFK} (b).
  \end{enumerate}
  \item The interactive measurement stage\\
        For each column x = 1,...,m in the brickwork state
        
        $ $ For each row y = 1,...,n in the brickwork state
  \begin{enumerate}
    \item Alice computes $\phi'_{x,y}$ based on the real measurement angle $\phi_{x,y}$ and the previous measurement results in order to achieve execution of the matching step of the MBQC (e.g. the gate in the application algorithm)
    \item Alice chooses a random bit $r_{x,y} \in \{0,1\}$ and sends the angle $\delta_{x,y} = \phi'_{x,y}+\theta_{x,y} + \pi r_{x,y}$, where $r$ is a random bit, to Bob via the classical channel.
    \item Bob measures $|\psi_{x,y}\rangle$ in the basis $\{|0\rangle \pm e^{i\delta_{x,y}}|1\rangle\}$ and sends the one-bit result to Alice via the classical channel.
    \item If $r_{x,y}=1$ above, Alice flips the result bit; otherwise she does nothing
  \end{enumerate}
\end{enumerate}

\subsection{Steane Code and Fault-tolerant Quantum Computation}
%Steane code
%prepare 0
%the crucial element of quantum error correction for stabilizer codes is the realization of the (parity) check measurement
%This idea of repeated code concatenation was used in the early days of quantum error correction to prove the Threshold Theorem which says that fault-tolerant computation is possible with arbitrary small error rate if one is willing to tolerate an overhead which scales polylogarithmically with the size N of the computation to be performed (the size of a quantum circuit is the number of locations in it).
%the error correction protocol cannot allow us to gain information regarding the coefficients, and of the encoded state
There are many different kinds of quantum error correction codes \cite{Gottesman09,DNM11,Terhal13}. In this paper, we will use a quantum error correction code called Steane's [[7,1,3]] code, which uses seven entangled physical qubits to encode one logical qubit and has the ability to correct one physical qubit error \cite{Steane96}. The [[7,1,3]] code is a stabilizer code, which is generally described by the stabilizer formalism \cite{Gottesman97}. A state $|\psi\rangle$ is stabilized by $K$ if $K|\psi\rangle = |\psi\rangle$. Every state in the [[7,1,3]] code is stabilized by these six operators,
\begin{equation}
\begin{aligned}
&K^{1}=IIIXXX, K^{2}=XIXIXIX, K^{3}=IXXIIXX,\\
&K^{4}=IIIZZZ, K^{5}=ZIZIZIZ, K^{6}=IZZIIZZ.
\end{aligned}
\end{equation}
Applying any of those operators to the state has no effect on the encoded logical state. Measuring whether the state is in the +1 or -1 eigenstate of each of these operators, as described below, gives us error syndrome information. These operators are useful in the synthesis of quantum circuits for preparing encoded states and error correction. To do a fault-tolerant quantum computation, we need to perform fault-tolerant quantum gates and fault-tolerant measurements on the error-correcting code blocks. A fault-tolerant quantum gate or a fault-tolerant measurement is implemented by a quantum circuit. Any single error in the circuit should not generate two error qubits in the same error-correcting code block. Thus, every error correction code block can correct its single qubit error after the fault-tolerant quantum gates or fault-tolerant measurements \cite{NC10,DNM11}.

\begin{figure}
\centerline{\scalebox{.4}{\includegraphics{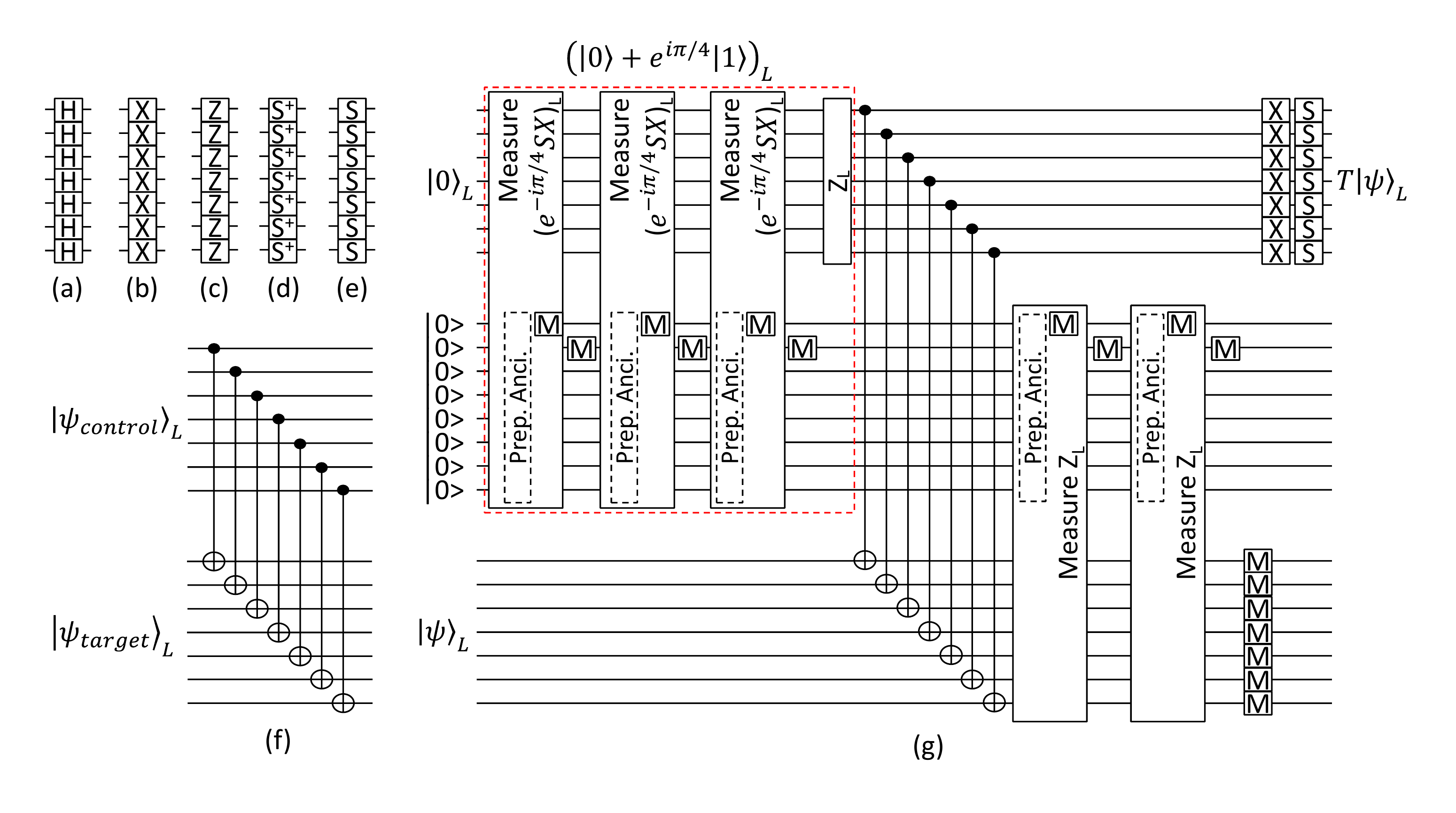}}}
\caption{Quantum circuit of fault-tolerant gates for the [[7,1,3]] code. (a) fault-tolerant Hadamard gate (b) fault-tolerant Pauli-X gate (c) fault-tolerant Pauli-Z gate (d) fault-tolerant S gate (e) fault-tolerant inverse S gate (f) fault-tolerant CNOT gate (g) fault-tolerant T gate.}
\label{fig:FTG}
\end{figure}

Any logical qubit $|\psi\rangle$ encoded by the  [[7,1,3]] code will be represented by $|\psi\rangle_{L}$. Any fault-tolerant gate $U$ for the [[7,1,3]] code will be represented by $U_{L}$. There are two kinds of fault-tolerant gates for the [[7,1,3]] code. For Clifford group gates, including Identity gates, Hadamard gates, Pauli gates, S gates, and CNOT gates, fault-tolerant gates are implemented by applying gates transversally on each qubit in the code as shown in the left of Fig.~\ref{fig:FTG}. For non-Clifford group gates, including T gates, fault-tolerant gates are implemented with the help of additional qubits, called ancilla qubits. The implementation of a fault-tolerant T gate for the [[7,1,3]] code is shown in Fig.~\ref{fig:FTG} (g). The red dashed box is the preparation of a special ancilla logical qubit $(|0\rangle + e^{i\pi/4}|1\rangle)_{L}$. If the result of fault-tolerant measuring the ancilla logical qubit $|0\rangle_{L}$ with operator $(e^{-i\pi/4}SX)_{L}$ is 1, a fault-tolerant $Z_{L}$ is applied to correct the output state. After applying $CNOT_{L}$ on this ancilla logical qubit and the input logical qubit, we fault-tolerantly measure the input logical qubit with operator $Z_{L}$ and then apply fault-tolerant SX on the ancilla logical qubit if the measurement result is 1. After that, the ancilla logical qubit is the output of this gate.

\begin{figure}
\centerline{\scalebox{.4}{\includegraphics{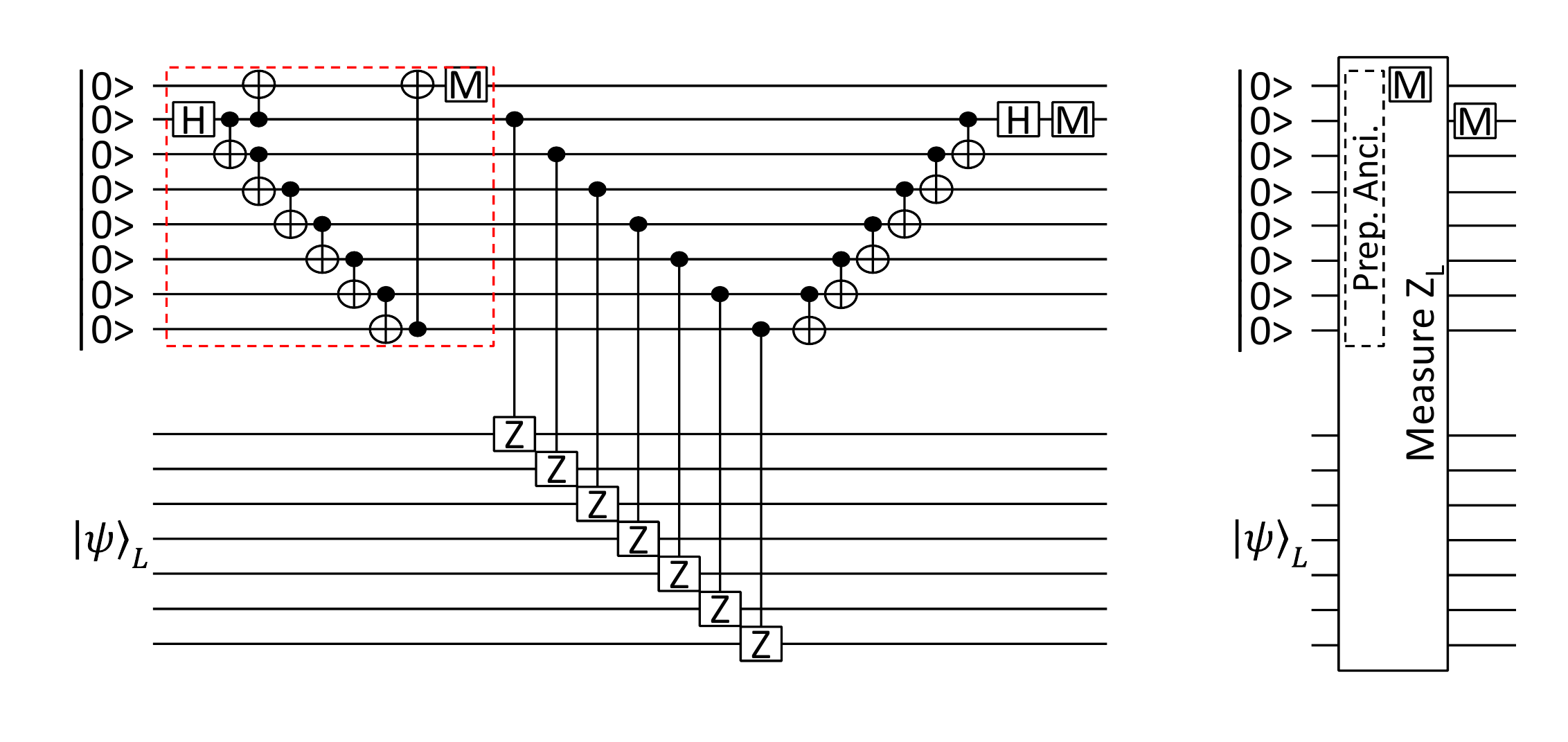}}}
\caption{Fault-tolerantly measuring the operator $Z_{L }$ on the [[7,1,3]] code}
\label{fig:FTMZ}
\end{figure}

\begin{figure}
\centerline{\scalebox{.4}{\includegraphics{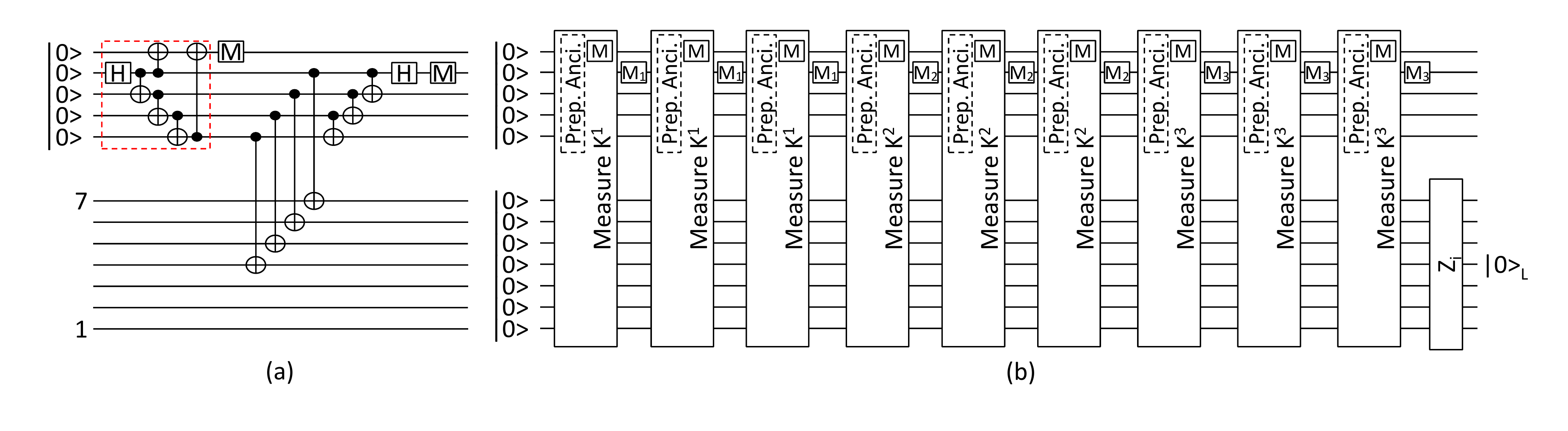}}}
\caption{(a) Fault-tolerantly measuring the operator $K^{1}$ on the [[7,1,3]] code. (b) Fault-tolerant preparation of a $|0\rangle_{L}$ state encoded with the [[7,1,3]] code.}
\label{fig:PrepZ}
\end{figure}

\begin{figure}
\centerline{\scalebox{.4}{\includegraphics{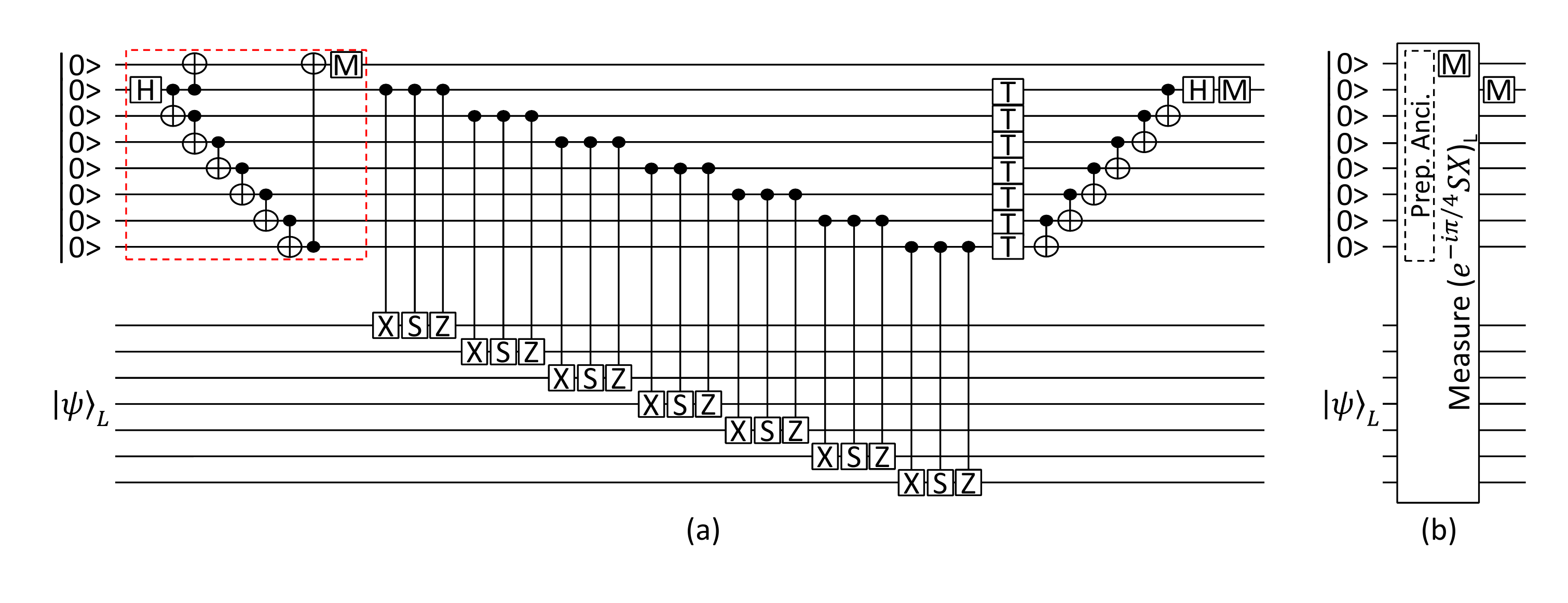}}}
\caption{Fault-tolerantly measuring the operator $(|0\rangle + e^{i\pi/4}|1\rangle)_{L}$ on the [[7,1,3]] code.}
\label{fig:PrepM}
\end{figure}

\begin{figure}
\centerline{\scalebox{.4}{\includegraphics{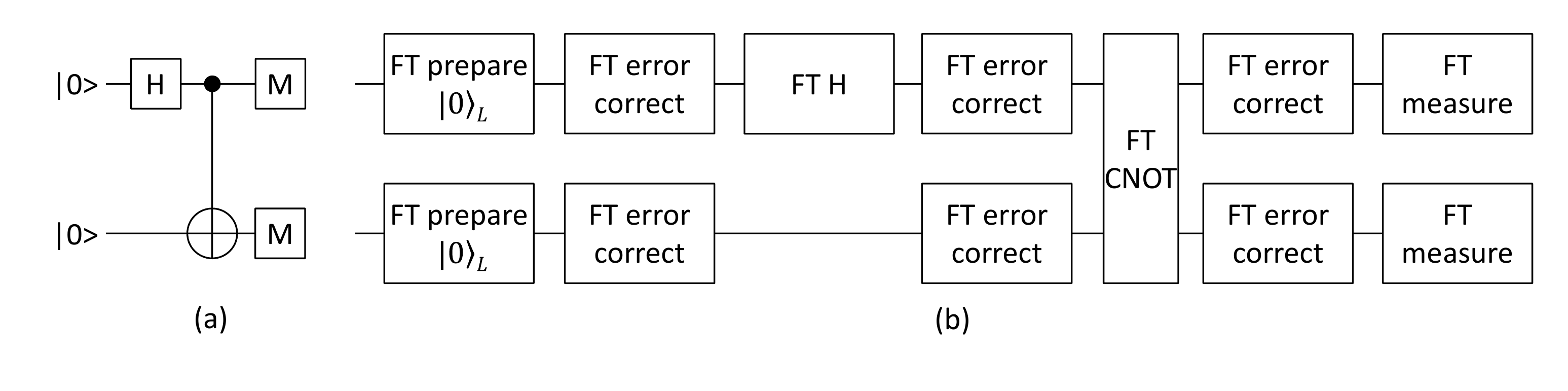}}}
\caption{(a) Sample quantum circuit (b) The fault-tolerant quantum circuit of the sample quantum circuit.}
\label{fig:FTQC}
\end{figure}

To measure an error correction code state fault-tolerantly, we do not measure them directly. Instead, we use ancilla qubits to extract information from the code state and then measure the ancilla qubits. The measurement result is a majority vote from three repetitions of this measurement procedure. Fig.~\ref{fig:FTMZ} shows a procedure for fault-tolerantly measuring operator $Z_{L}$ on the [[7,1,3]] code. The red dashed box shows seven ancilla qubits are prepared as the state $\frac{1}{\sqrt{2}}(|0000000\rangle + |1111111\rangle)$ and they are verified by measuring another ancilla qubit.
Fault-tolerant measurements are also used to prepare each encoded state. The $|0\rangle_{L}$ state is prepared by fault-tolerantly measuring seven $|0\rangle$ qubits with operator $K^{1}, K^{2}$, and $K^{3}$. Depending on the measurement results, a $Z$ gate may be applied to one of the qubits in the code block to make the correct $|0\rangle_{L}$ state. Fig.~\ref{fig:PrepZ} shows the fault-tolerant preparation of a $|0\rangle_{L}$ state encoded with the [[7,1,3]] code.
By applying additional fault-tolerant quantum gates, any other code state can be prepared. The special state $(|0\rangle + e^{i\pi/4}|1\rangle)_{L}$ used in the fault-tolerant T gate is prepared by fault-tolerantly measuring operator $e^{-i\pi/4}SX$ on state $|0\rangle_{L}$ as shown in Fig.~\ref{fig:PrepM}.
Quantum error correction is achieved by extracting the error syndrome from the code state and then applying corrections to the code state. The error syndrome extraction for the [[7,1,3]] code is also performed by fault-tolerantly measuring the code state with the operators $K^{1}, K^{2}, K^{3}, K^{4}, K^{5}$, and $K^{6}$. Depending on the measurement results, a $Z$ gate and a $X$ gate may be applied to one of the qubits in the code to correct the error. Fig.~\ref{fig:FTQC} shows an example of how to build a fault-tolerant quantum circuit with quantum error correction \cite{NC10}.

%According to the error syndrome, we can correct the error by applying additional quantum gates.  Fault-tolerant measurement are also used to prepare encoded state. 

\subsection{BFK Fault-tolerant Blind Quantum Computation}
Broadbent {\em et al.} showed that their blind quantum computation can be made fault-tolerant \cite{BFK09}. Alice first converts the target computation to a fault-tolerant circuit. The first phase of the circuit is building the quantum state for the input variables, encoding the input data into a quantum error correction code. Within the code block, all qubit wires are permuted. For the non-Clifford group fault-tolerant gates and fault-tolerant measurement, ancilla qubits are required and they should be measured in the computational basis. Ancilla qubit wires are added and evenly spaced through the circuit and they are re-used. All the ancilla qubits have to be measured at the same time, at a regular interval, after each fault-tolerant gate. Since some of the ancilla are not used at the time of measurement, their measurement results are meaningless. Then Alice converts the fault-tolerant circuit to a measurement-based computation on the brickwork state. Ancilla qubits are still measured in the computational basis, which is different from other measurements. Lastly, Alice does this fault-tolerant computation using the basic blind quantum computation. In addition, she has to periodically instruct Bob to measure all ancilla qubits in the computational basis.

In the FT form of BFK, only Bob does quantum error correction but not Alice. Error correction is run \emph{on top of the brickwork state}, requiring Alice to directly process all of the phases of error correction, including syndrome extraction and correction operations. This approach has two drawbacks: first, the amount of classical communication required between Alice and Bob is enormous, and classical communication latencies may become important; second, because the individual underlying qubits are not error-protected, the error rate induced by the initial teleportation stage may become more of an issue.

%PLUS: Require Bob more quantum computation power
%the preparation of the input state is included in the computational part.
\section{Blind Quantum Computation Using Fault-tolerant Circuits}
% MBQC supports long-distance gates in a single time step even when the cluster state is built on a physical system permitting only nearest-neighbor interactions.
To improve the fault-tolerance of blind quantum computation, we propose the first fault-tolerant quantum computation protocol which applies quantum error correction underneath blind quantum computation. All of the qubits in the blind quantum computation are encoded by one layer of the [[7,1,3]] code and all quantum operations are replaced by fault-tolerant operations. Thus, Alice's computational capability must grow to have the ability to fault-tolerantly prepare error correction code blocks. Bob will use a fault-tolerant circuit to implement measurement-based quantum computation in blind quantum computation. 

In our first fault-tolerant protocol, Alice has to prepare input logical qubits encoded in one layer of the [[7,1,3]] code. All logical qubits are prepared by applying fault-tolerant phase shift gates to the $\frac{1}{\sqrt{2}}(|0\rangle+|1\rangle)_{L}$ code state. $\frac{1}{\sqrt{2}}(|0\rangle+|1\rangle)_{L}$ is prepared by applying a fault-tolerant Hadamard gate to initial $|0\rangle_{L}$ code state. To prepare $\frac{1}{\sqrt{2}}(|0\rangle + e^{i\theta}|1\rangle)_{L}$, where $\theta = 0,\pi/4,2\pi/4,3\pi/4,4\pi/4,5\pi/4,6\pi/4,7\pi/4$, Alice applies $I_{L}$, $T_{L}$, $S_{L}$, $(TS)_{L}$, $Z_{L}$, $(TZ)_{L}$, $(S^{\dagger})_{L}$, $(TS^{\dagger})_{L}$ respectively to the $\frac{1}{\sqrt{2}}(|0\rangle+|1\rangle)_{L}$ code state. Thus, Alice is required to do a small fault-tolerant quantum computation to prepare each encoded logical qubit.

Bob also has to change his operations in our first fault-tolerant blind quantum computation protocol. His operation is similar to the BFK basic blind quantum computation protocol, except all quantum operations are fault-tolerant. In the basic protocol, Bob has to do two things: first, he entangles the received qubits to create a brickwork state; second, he sequentially measures each qubit in the $M(\delta_{x,y})$ basis, where ${x,y}$ represents the qubit in the $x$ column and $y$ row of the brickwork state, using the basis requested by Alice, and tells Alice the measurement results. The brickwork state can be divided into layers and then be divided into bricks as shown in Fig.~\ref{fig:BFK} (b). The output qubits of each layer are the input qubits of the following layer. After the blind quantum computation is finished in the first layer, the second layer starts, and so on. Within each layer, every brick is independent. Thus, if Bob can do the fault-tolerant blind quantum computation on a brick, he can do the fault-tolerant blind quantum computation on the brickwork state.

\begin{figure}
\centerline{\scalebox{.4}{\includegraphics{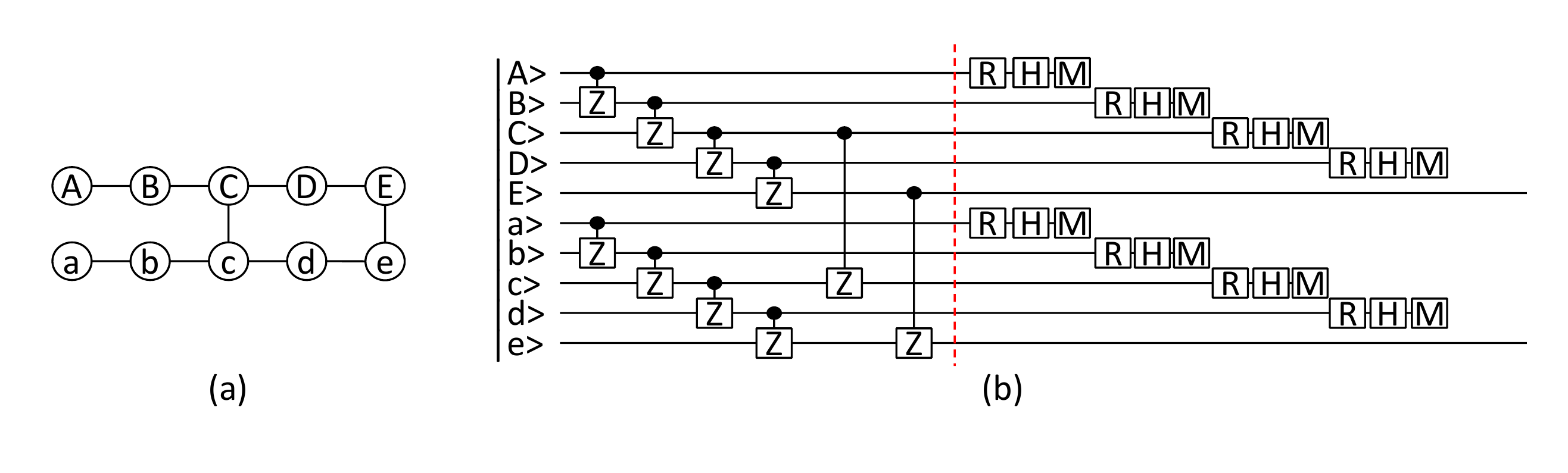}}}
\caption{The equivalent quantum circuit for the server's quantum operation on a brick in blind quantum computation, including entanglements, phase shift gates, Hadamard gates, and measurements. R represents a phase shift operation, which can be implemented by a composition of Z gate, S gate, and T gate. M represents measurement in the computational basis.}
\label{fig:brickwork}
\end{figure}

Bob's quantum operations on a brick in the blind quantum computation are implemented by an equivalent quantum circuit as shown in Fig.~\ref{fig:brickwork} (b). Ten qubits are entangled by CZ gates, then eight qubits are measured in the $M(\delta_{x,y})$ basis. For measurement basis $M(\delta_{x,y})$, where $\delta_{x,y} = 0, \pi/4, 2\pi/4, 3\pi/4, 4\pi/4, 5\pi/4, 6\pi/4, 7\pi/4$, Bob applies $H$, $HTS^{\dagger}$, $HS^{\dagger}$, $HTZ$, $HZ$, $HTS$, $HS$, or $HT$, respectively, before a measurement in the computational basis. The phase shift gates, Hadamard gates, and measurement should be performed in the order of alphabet indexed in the figure. Then this quantum circuit is transformed to a fault-tolerant quantum circuit which uses one layer of the [[7,1,3]] code. Each qubit is replaced by a seven-qubit code block and all gates are changed to fault-tolerant quantum gates. For the fault-tolerant T gates and the fault-tolerant measurement in the computational basis, ancilla qubits are added to the circuit. Since the quantum circuit of blind quantum computation on a brick is fault-tolerant, the blind quantum computation on a brickwork state by repeating this quantum circuit is fault-tolerant.

Our first fault-tolerant protocol is shown in Protocol ~\ref{ptl:first}. Alice prepares each qubit in the brickwork state by fault-tolerantly preparing encoded $|0\rangle_{L}$ state and fault-tolerantly applying a random phase shift on the code state. Then she sends it to Bob. Bob fault-tolerantly entangles all the encoded logical qubits to create a brickwork state using fault-tolerant CZ gates as shown in Fig.~\ref{fig:brickwork} (b). Then Bob performs a fault-tolerant phase shift $(-\delta_{x,y})_{L}$, which is assigned by Alice, on each encoded logical qubit $|\psi_{x,y}\rangle_{L}$ before measuring it in the computational basis. The phase shifts assigned by Alice are still calculated as in the BFK basic protocol. The phase shifts and measurements still follow the order of the column of the qubits. Since every qubit in this protocol is encoded with the error correction code and all quantum operations are performed by fault-tolerant quantum circuit, this protocol is fault-tolerant. Except for the underlying quantum error correction code and the fault-tolerant quantum circuit, this protocol follows the BFK basic blind quantum computation. The fault-tolerant circuit to prepare encoded logical qubits, to entangle encoded logical qubits, and to measure encoded logical qubits are all independent of the inputs, outputs and the quantum gates in the target computation. Thus, the blindness of this protocol is the same as the BFK basic blind quantum computation. This protocol reveals only the information of the upper bound of the size of the computation.
% Algorithm
\renewcommand{\algorithmcfname}{PROTOCOL}
\begin{algorithm}[t]
\caption{Fault-tolerant Blind Quantum Computation Using Fault-tolerant Circuit}
\label{ptl:first}
\SetAlgoNoLine
\begin{enumerate}
  \item The preparation stage
  \begin{enumerate}
    \item For each encoded logical qubit $|\psi_{x,y}\rangle_{L}$ in the brickwork state, Alice fault-tolerantly prepares a $\frac{1}{\sqrt{2}}(|0\rangle+|1\rangle)_{L}$ logical qubit encoded in one layer of the Steane [[7,1,3]] code and applies a fault-tolerant phase shift $\theta_{x,y}$, where $\theta_{x,y}$ is randomly chosen from $\{0, \pi/4, 2\pi/4, 3\pi/4, 4\pi/4, 5\pi/4, 6\pi/4, 7\pi/4\}$. Then the encoded logical qubits are sent to Bob via the quantum channel.
    \item Bob fault-tolerantly entangles all the encoded logical qubits to create the brickwork state using fault-tolerant fault-tolerant CZ gates.
  \end{enumerate}
  \item The interactive measurement stage\\
        For each column x = 1,...,m in the brickwork state
        
        $ $ For each row y = 1,...,n in the brickwork state
  \begin{enumerate}
    \item Alice computes $\phi'_{x,y}$ based on the real measurement angle $\phi_{x,y}$ and the previous measurement results. 
    \item Alice chooses a random bit $r_{x,y} \in \{0,1\}$ and sends $\delta_{x,y} = \phi'_{x,y}+\theta_{x,y} + \pi r_{x,y}$ to Bob via the classical channel.
    \item Bob applies a fault-tolerant phase shift $(-\delta_{x,y})_{L}$ and a fault-tolerant Hadamard gate on the encoded logical qubit $|\psi_{x,y}\rangle_{L}$ and fault-tolerantly measures the encoded logical qubit in the computational basis. The one-bit measurement result is sent to Alice via the classical channel.
    \item If $r_{x,y}=1$ above, Alice flips the result bit; otherwise she does nothing.
  \end{enumerate}
\end{enumerate}
\end{algorithm}

\begin{figure}
\centerline{\scalebox{.3}{\includegraphics{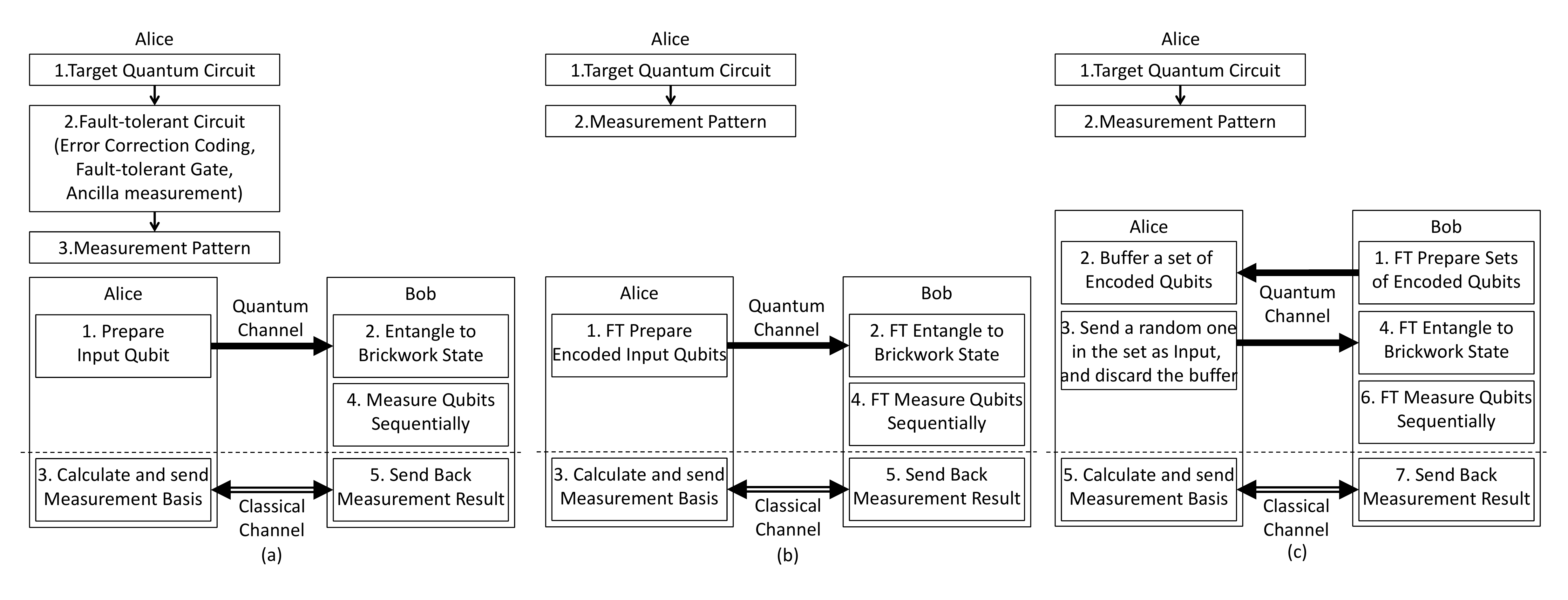}}}
\caption{Comparison of fault-tolerant blind quantum computation. The upper part shows Alice's preparation before the fault-tolerant blind quantum computation. The lower part shows the steps of protocol. The dashed line separates the quantum part and the classical part of the blind quantum computation. (a) The BFK fault-tolerant protocol (b) Our first fault-tolerant protocol (c) Our second fault-tolerant protcol, the BSA protocol.}
\label{fig:FTBQCcomp}
\end{figure}

From Fig.~\ref{fig:FTBQCcomp}, we can clearly see the difference between the BFK fault-tolerant protocol and our first fault-tolerant protocol using fault-tolerant circuits. The top part of the figure shows what Alice should prepare before starting the blind quantum computation protocol. The bottom part of the figure shows how the blind quantum computation proceeds. In the BFK fault-tolerant protocol, Alice has to convert her target computation to a fault-tolerant quantum circuit with additional evenly-separated ancilla qubits measured in the computational basis at the same time with a regular interval. Then she converts the fault-tolerant quantum circuit to measurement patterns for the blind quantum computation. After that, she can perform the blind quantum computation using the basic protocol, with additional periodic measurements in the computational basis. In our first fault-tolerant protocol using fault-tolerant circuits, Alice converts her target circuit to measurement patterns for the blind quantum computation before she starts the fault-tolerant blind quantum computation protocol. Alice has to do fault-tolerant quantum computation to prepare the encoded input qubits. Bob fault-tolerantly entangles all the qubits and fault-tolerantly measures them.

Compared to the BFK fault-tolerant protocol, our first fault-tolerant blind quantum computation has two advantages. First, our first fault-tolerant protocol provides more fault-tolerance. Teleporting QEC-encoded states protects against errors in our qubits that occur during the teleportation process. At the physical level, quantum repeater networks may reach reasonably high fidelities, but the error rates necessary to execute e.g. $10^{15}$ logical gates successfully are out of the question. Our approach will be useful if Alice and Bob are not directly connected. Second, Alice prepares fewer physical qubits and the amount of classical communication required between Alice and Bob is far less. In our first fault-tolerant protocol, Alice prepares a constant factor (seven times for Steane's [[7,1,3]] code) more qubits than she needs in the BFK basic protocol. In the BFK fault-tolerant protocol, performing the fault-tolerant quantum computation on top of the brickwork state makes the brickwork state a linear growth because the number of additional swap gates is linear in the size of the computation. These additional swap gates are used to avoid CNOT gates operating on non-adjacent qubits. Thus, Alice has to prepare a linear factor more qubits than she needs in the BFK basic protocol. Since Alice and Bob need to exchange classical messages between each measurement, preparing and measuring more qubits requires stricter conditions on classical communication latencies.

Our first fault-tolerant protocol has two disadvantages. First, Alice is required to do fault-tolerant quantum computation to prepare encoded input qubits, and her quantum computational overhead is proportional to the qubits she prepared, which is the total size of the brickwork state. This is a tough requirement for Alice, who has limited quantum computing power. Second, Alice may leak some information which can be exploited by a side-channel attack. Since the fault-tolerant T gates consumes much more quantum resources than other fault-tolerant gates, preparing encoded logical qubits $\frac{1}{\sqrt{2}}(|0\rangle + e^{i\theta}|1\rangle)_{L}$, where $\theta = \pi/4, 3\pi/4, 5\pi/4, 7\pi/4$, consumes more power from Alice. If Bob or a third-party attacker can probe the computing power or the preparation time during Alice's preparation of encoded logical qubits, they can learn if the encoded logical qubits are prepared by fault-tolerant T gates. Thus, this protocol will not be perfectly secure. To overcome these two drawbacks of our first fault-tolerant protocol, we propose our second fault-tolerant protocol which requires Alice only to receive, buffer and send qubits.

%\section{New Approach of Blind Quantum Computation Using Fault-tolerant Circuit}
%Compare to other
\section{Buffer Shuffle Alice Protocol}
To lower Alice's requirement of quantum computing power and reduce Alice's computation overhead, we propose the second protocol, which lets Bob prepare the encoded logical qubits. This protocol is called the Buffer Shuffle Alice (BSA) Protocol.  Since Alice cannot reveal any input information to Bob, Alice must retain the decision of selecting input logical qubits. The second fault-tolerant blind quantum computation protocol also has two stages. The first stage is the preparation stage. A brickwork state will be created for the blind quantum computation. For each encoded logical qubit in the brickwork state, Bob fault-tolerantly prepares a fixed set of eight different logical qubits encoded in the [[7,1,3]] code and sends them in a fixed order to Alice via the quantum channel. Alice has to buffer these encoded logical qubits before she sends back to Bob an encoded input qubit which is one encoded logical qubit randomly chosen from these encoded logical qubits. Then Bob entangles all received encoded logical qubits to create a brickwork state by performing a fault-tolerant quantum circuit as shown in Fig.~\ref{fig:brickwork} (b). The second stage is the interactive measurement stage, which is the same as our first protocol. Alice calculates the $\delta$ of the measurement basis $M(\delta)$ and sends the value to Bob via the classical channel. Bob fault-tolerantly measures each encoded logical qubits in the $M(\delta)$ basis and sends back the measurement result to Alice for her later calculation. This protocol is shown in Protocol ~\ref{ptl:second}.
\renewcommand{\algorithmcfname}{PROTOCOL}
\begin{algorithm}[t]
\caption{Buffer Shuffle Alice Protocol}
\label{ptl:second}
\SetAlgoNoLine
\begin{enumerate}
  \item The preparation stage
  \begin{enumerate}
    \item For each encoded logical qubit $|\psi_{x,y}\rangle_{L}$ in the brickwork state, Bob fault-tolerantly prepares a set of eight $\frac{1}{\sqrt{2}}(|0\rangle+|1\rangle)_{L}$ logical qubits encoded in one-layer of the Steane's [[7,1,3]] code and applies fault-tolerant phase shift gates $\theta_{x,y}$, where $\theta_{x,y} = 0, \pi/4, 2\pi/4, 3\pi/4, 4\pi/4, 5\pi/4, 6\pi/4, 7\pi/4$ to these qubits respectively. Then the set of encoded logical qubits is sent to Alice in a fixed order via the quantum channel.
    \item Alice buffers the received logical qubits before she receives a set of encoded logical qubits. She sends back one encoded logical qubits randomly chosen from the set as an input logical qubit and discards the rest.
    \item Bob fault-tolerantly entangles all the encoded logical qubits to create the brickwork state using fault-tolerant CZ gates.
  \end{enumerate}
  \item The interactive measurement stage\\
        For each column x = 1,...,m in the brickwork state
        
        $ $ For each row y = 1,...,n in the brickwork state
  \begin{enumerate}
    \item Alice computes $\phi'_{x,y}$ based on the real measurement angle $\phi_{x,y}$ and the previous measurement results. 
    \item Alice chooses a random bit $r_{x,y} \in \{0,1\}$ and sends $\delta_{x,y} = \phi'_{x,y}+\theta_{x,y} + \pi r_{x,y}$ to Bob via the classical channel.
    \item Bob applies a fault-tolerant phase shift $(-\delta_{x,y})_{L}$ and a Hadamard gate on the encoded logical qubit $|\psi_{x,y}\rangle_{L}$ and fault-tolerantly measures the encoded logical qubit in the computational basis. The one-bit measurement result is sent to Alice via the classical channel.
    \item If $r_{x,y}=1$ above, Alice flips the result bit; otherwise she does nothing.
  \end{enumerate}
\end{enumerate}
\end{algorithm}

Since every qubit in the BSA protocol is encoded with quantum error correction code and all quantum operations are performed by fault-tolerant quantum circuits, this protocol is fault-tolerant. The underlying error correction code and the fault-tolerant quantum circuit in the BSA protocol is independent of the input, output, and quantum gates in the target computation. The BSA protocol follows the BFK basic blind quantum computation protocol except the preparation of the input qubits for the brickwork state. Bob prepares a fixed set of eight qubits with different phases and sends them in a fixed order, which is known by Alice and Bob. Alice randomly chooses one encoded logical qubit as an input logical qubit. The phase of the encoded logical qubit is a random number and is only known by Alice. Since Alice prepares no qubits and only tells Bob the classical information of the measurement basis for each qubit in the brickwork state. We only need to prove that this classical information reveals no information of the quantum computation. Let $\theta$ be the set of chosen phases of every encoded input qubit. Let $\phi'$ be the set of modified angles of measurement basis for the implementation of the quantum gate depending on the previous measurement results. Let $r$ be a random bit string generated by Alice. The angle of the measurement basis which Alice tells Bob is calculated as follows.
\begin{equation}
\delta_{x,y} = \phi'_{x,y} + \theta_{x,y} + \pi r_{x,y},
\end{equation}
where $x,y$ represents the qubit is in the $x$ column and $y$ row of the brickwork state. Since $\theta$ is a random number, $\delta$ is independent of $\phi'$. The angles of measurement basis which Alice tells Bob reveal no information of the quantum gates in the blind quantum computation. Since $r$ is a random bit, $\delta$ is also independent of $\theta$. The angles of measurement basis which Alice tells Bob reveals no information of the input and output qubits in the blind quantum computation, too. Thus, we have shown that Alice reveals no information of the target quantum computation in this protocol, except the number of input qubits, which is an upper bound of the size of the computation.

%Possible attack
Compared to our first fault-tolerant protocol, the BSA protocol avoids the additional quantum computing requirement for Alice and still preserves the same fault-tolerance ability. Since Alice doesn't have to do fault-tolerant quantum computation to prepare different encoded logical qubits, the side-channel attack described in the last section will not work on the BSA protocol. The BSA protocol has two differences from the BFK fault-tolerant protocol, see Fig.~\ref{fig:FTBQCcomp}. First, quantum error correction and fault-tolerant quantum circuit is applied \emph{underneath} the blind quantum computation in the BSA protocol, while quantum error correction and fault-tolerant quantum circuit is applied \emph{on top of} the blind quantum computation in the BFK fault-tolerant protocol. Second, Alice buffers and selects encoded logical qubits prepared by Bob in the BSA protocol, while Alice prepares qubits with random phase shifts in the BFK fault-tolerant protocol.

Compared to the BFK fault-tolerant protocol, the BSA protocol has the following advantages. First, the demands made of Alice's system are very low. She only buffers, receives, and sends encoded logical qubits. Second, the BSA protocol provides more fault-tolerance. Alice and Bob can be connected by a long and noisy channel if quantum error correction can be performed on the states end-to-end. Third, the amount of quantum and classical communication required between Alice and Bob is less. For each logical qubit in the brickwork state, Bob sends eight logical qubits, which are 56 physical qubits, to Alice. Alice returns one logical qubit to Bob. The total number of physical qubits transmitted in the quantum channel is 63 times the number of logical qubits in the brickwork state. In BFK fault-tolerant protocol, the size of the brickwork state will be increased by a linear factor compared to the computational size because additional swap gates are used to avoid CNOT gates operating on non-adjacent qubits.

There are two trade-off in the BSA protocol. First, Alice has to buffer eight logical qubits at a time, a total of 56 physical qubits if the Steane's [[7,1,3]] code is used. But she only needs to buffer eight receiving logical qubits before she selects and sends one of them. The buffering time depends on the quantum transmission rate and the quantum error correction code. Second, the BSA protocol consumes nine times as much network bandwidth (that is, nine times as many end-to-end Bell pairs are requested from the repeater network) as simply having Alice prepare and teleport the logical qubits in our first fault-tolerant protocol. The total memory buffering requirement for Alice can actually be reduced to two logical qubits, if Alice chooses to measure and discard all but one of the eight qubits as they arrive. More complex buffering and shuffling schemes could be used to reduce the 9x bandwidth penalty. We defer detailed analysis of this to future work.

From the BFK fault-tolerant protocol and our two fault-tolerant protocols, we can see that fault-tolerant blind quantum computation costs much more quantum resources than the basic BFK blind quantum computation protocol. In the BFK fault-tolerant protocol, the more complex fault-tolerant quantum circuit requires a larger size of the brickwork state. Alice also has to prepare more qubits. In our two fault-tolerant protocols, transmitting encoded logical qubits via quantum channel consumes more quantum network bandwidth. In our first fault-tolerant protocol, Alice even has to do quantum computation to prepare encoded logical qubits. We will examine the resource consumption in our two fault-tolerant protocols in the next section.

% Head 2
\section{resource analysis}
\begin{figure}
\centerline{\scalebox{.6}{\includegraphics{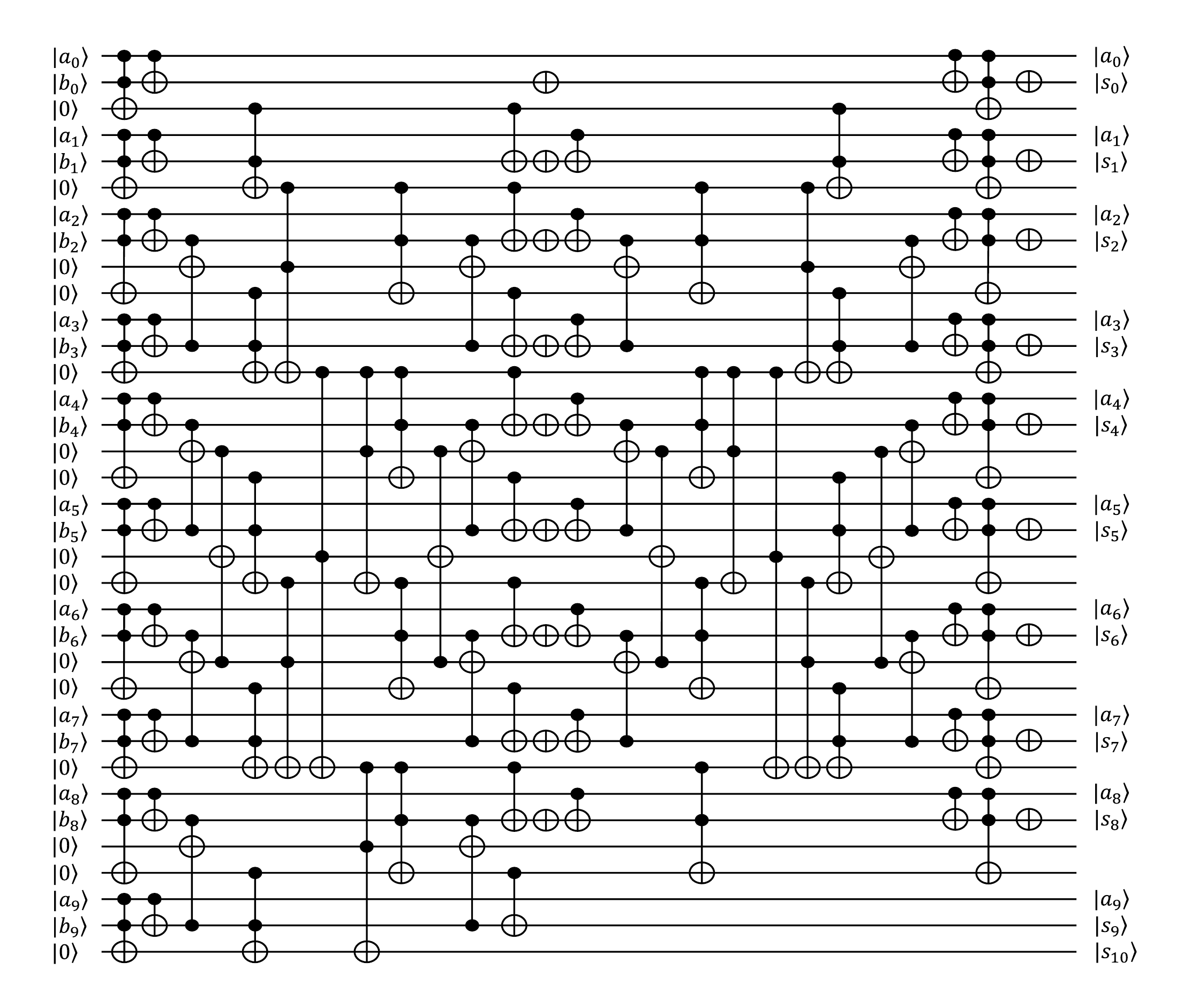}}}
\caption{An 10-bit in-place quantum carry-lookahead adder. $s$ is the sum of $a$ and $b$. The subscript indexes the bit in the number.  \cite{DKRS06}}
\label{fig:QCLA10}
\end{figure}

%QCLA adder can be used within current modular multiplication circuits to reduce substantially the run-time of shor's algorithm.
%Addition is a critical subroutine for algorithms such as Shor's algorithm for factoring large numbers
%depth linear
To show the difference in resource consumption of our protocols, we analyse the blind quantum computation of a 10-bit quantum carry-lookahead (QCLA) adder \cite{DKRS06}. A quantum adder is a useful component for constructing multipliers and more elaborate circuits. It is also a basic subroutine in Shor's algorithm \cite{Shor97,VBE96,BCDP96}. The QCLA adder has a logarithmic depth in the number of input qubits. A quantum circuit for a 10-bit in-place QCLA adder is illustrated in Fig.~\ref{fig:QCLA10}.

To do a blind quantum computation, we need to convert the target quantum circuit to measurement-based computation on a brickwork state. First, the quantum circuit is converted to an equivalent circuit using single-qubit gates and CNOT gates. Second, we add swap gates in the circuit to make sure all CNOT gates operate on adjacent qubits. Third, we arrange each quantum gate to fit in a brick of the brickwork state. Each line in the circuit corresponds to a row of qubits in the brickwork state. Since the bricks are even-odd interleaved in the brickwork state, CNOT gates can only be arranged in specific layers. Fourth, in each brick, we assign each qubit a measurement basis according to the quantum gate implemented by the brick.

\begin{figure}
\centerline{\scalebox{.5}{\includegraphics{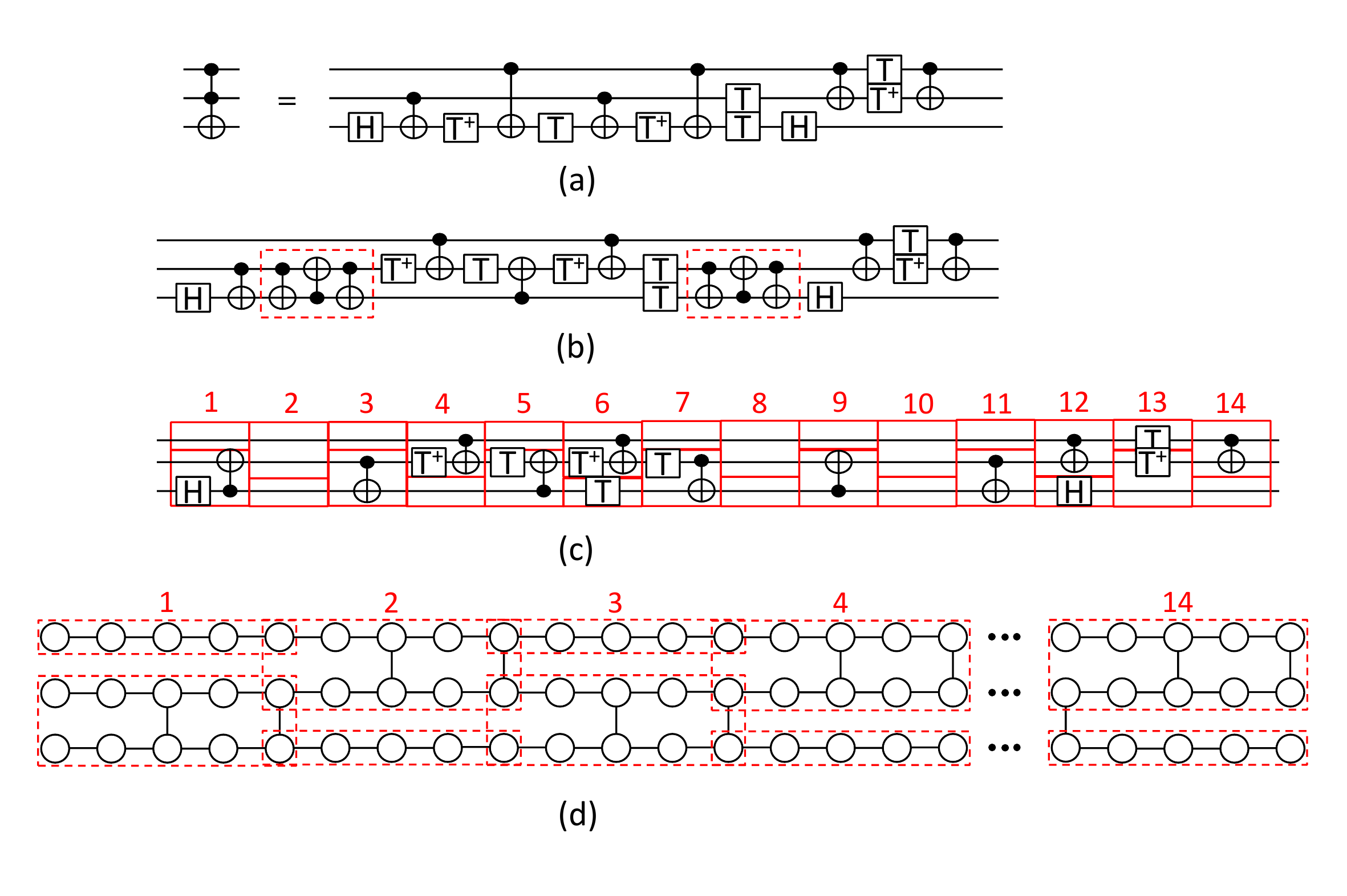}}}
\caption{(a) Decomposition of a Toffoli gate. (b) Decomposed circuit where CNOT gates only operate on adjacent qubits. Dashed boxes represent SWAP gates, which are decomposed to three consecutive CNOT gates. (c) Arrange quantum gates to fit in the bricks in the brickwork state. (d) The brickwork state required to implement a Toffoli gate.}
\label{fig:decomp}
\end{figure}

We will take a Toffoli gate as an example to show how to convert the quantum circuit to measurement-based quantum computation. First, a Toffoli gate is decomposed to one-qubit gates and two-qubit gates as shown in Fig.~\ref{fig:decomp} (a) \cite{SM09}. Second, two swap gates are added to make CNOT gates operate on adjacent qubits as shown in Fig.~\ref{fig:decomp} (b). Swap gates are also decomposed to three consecutive CNOT gates, as there is no simpler construction of swap on the brickwork state. Third, all quantum gates are arranged into a brick in the brickwork state as shown in Fig.~\ref{fig:decomp} (c). Some one-qubit gates and the following CNOT gates can be implemented in the same brick. Fig.~\ref{fig:decomp} (d) shows the brickwork state required to do the decomposed circuit of a Toffoli gate in blind quantum computation. There are 14 layers of bricks. Three input qubits in the circuit makes three rows of qubits in the brickwork state. Thus, this brickwork state consists of 171 qubits. A brick consists of 10 qubits, but some qubits in the first row and in the last row cannot form a brick. The entangled five-qubit groups, called half-bricks, can also implement one-qubit gates. The last step is to assign each qubit a measurement basis according to the quantum gate implemented by the brick. For those half-bricks or bricks which implement no quantum gates in the circuit, they just implement one or two single-qubit identity gates. Thus every qubit is assigned a measurement basis, except the last column of qubits which are the output qubits.

Before calculating the resource consumption of a 10-bit QCLA adder in our protocol, we first calculate the resource consumption of the fault-tolerant quantum circuit for a 10-bit QCLA adder. The original circuit of a 10-bit QCLA adder has 35 input qubits, and it consists of 63 Toffoli gates, 35 CNOT gates, and 18 NOT gates. Each Toffoli gate can be decomposed into 7 $T/T^{\dagger}$ gates, 6 $CNOT$ gates, and 2 $H$ gates. With the decomposition of the Toffoli gates, the decomposed circuit consists of 441 $T/T^{\dagger}$ gates, 413 $CNOT$ gates, and 144 one-qubit Clifford group gates, and the number of its input qubits remains the same. To implement a fault-tolerant T gate, we need to prepare a $|0\rangle_{L}$ state before preparing $\frac{1}{\sqrt{2}}(|0\rangle + e^{i\pi/4}|1\rangle)_{L}$ state. From Fig.~\ref{fig:PrepZ}, the fault-tolerant circuit to prepare $|0\rangle_{L}$ state consists of 108 CNOT gates, 18 Hadamard gates, and a Z gate. Form Fig.~\ref{fig:FTG} (g), a fault-tolerant $T$ gate consists of 21 $T$ gate, 262 two-qubit Clifford group gates, 50 one-qubit Clifford group gates, and 35 measurements. Other fault-tolerant Clifford-group gates are implemented by applying non-fault-tolerant gates transversally on each physical qubit. If we apply one layer of Steane's [[7,1,3]] code and fault-tolerant gates to this circuit, the fault-tolerant decomposed circuit consists of 9,261 $T/T^{\dagger}$, 118,433 two-qubit Clifford group gates, 23,058 one-qubit Clifford group gates, and 15,435 measurements. This fault-tolerant circuit requires 245 qubits and at least 15 ancilla qubits for fault-tolerant T gates.

In our first fault-tolerant protocol, Alice has to prepare the encoded logical qubits. We will calculate the resource consumption of this process. All qubits are prepared from an initial $|0\rangle_{L}$ state. To prepare $\frac{1}{\sqrt{2}}(|0\rangle + e^{i\theta}|1\rangle)_{L}$, where $\theta = 0, 2\pi/4, 4\pi/4, 6\pi/4$, Alice applies fault-tolerant $H_{L}$, $(SH)_{L}$, $(ZH)_{L}$, and $(S^{\dagger}H)_{L}$ to the $|0\rangle_{L}$ state. To prepare $\frac{1}{\sqrt{2}}(|0\rangle + e^{i\theta}|1\rangle)_{L}$, where $\theta = \pi/4, 3\pi/4, 5\pi/4, 7\pi/4$, Alice applies fault-tolerant $(TH)_{L}$, $(TSH)_{L}$, $(TZH)_{L}$, and $(TS^{\dagger}H)_{L}$ to the $|0\rangle_{L}$ state. On average, Alice uses 10.5 T gates, 239 two-qubit Clifford-group gates, 56.25 one-qubit Clifford-group gates, and 35.5 measurements in preparing each encoded logical qubit. Alice also needs to operate at least 22 physical qubits because a fault-tolerant T gate requires at least 15 ancilla qubits.

In our fault-tolerant blind quantum computation, for each brick, Bob will perform 10 fault-tolerant CZ gates, 8 fault-tolerant phase shift, 8 fault-tolerant Hadamard gates, and 8 fault-tolerant measurements in the computational basis. Bob has to do eight different fault-tolerant phase shift, which are implemented by $I_{L}$, $T_{L}$, $S_{L}$, $(TS)_{L}$, $Z_{L}$, $(TZ)_{L}$, $S^{\dagger}_{L}$, and $(TS^{\dagger})_{L}$. On average, a phase shift requires 10.5 T gates, 131 two-qubit Clifford-group gates, 30.25 one-qubit Clifford-group gates, and 17.5 measurements. Fault-tolerantly measuring an encoded logical qubit in the computational basis requires 42 CNOT gates, 4 Hadamard gates, and 11 measurements. Thus, for each brick, Bob has to perform 84 T gates, 1454 two-qubit Clifford-group gates, 330 one-qubit Clifford-group gates, and 228 measurements. For each half-brick, Bob has to perform 42 T gates, 720 two-qubit Clifford-group gates, 165 one-qubit Clifford-group gates, and 114 measurements.

Here we will calculate the resource consumption of performing a 10-bit QCLA adder on blind quantum computation. First, we know the decomposed circuit of a 10-bit QCLA adder consists of 441 $T/T^{\dagger}$ gates, 413 $CNOT$ gates, and 144 one-qubit Clifford group gates, and it has 35 input qubits. Second, we added 328 swap gates to the circuit to make CNOT gates operate on adjacent qubits. Third, after arranging quantum gates to bricks with hand-optimization, we need 612 layers of bricks. Since it has 35 input qubits, we need a brickwork state with 35 rows. Thus, the totoal number of qubits in this brickwork state is $(1+612\times4)\times35 = 85,715$. The brickwork state consists of 10,404 bricks and 612 half-bricks.

%no rotate
In the BFK basic blind quantum computation, Alice needs a special qubit generator, and she has to prepare 85,715 qubits with random phase. For each brick, Bob has to perform 10 entanglements, 8 phase shift, 8 Hadamard gates and 8 measurements in the computational basis. For each half-brick, Bob has to perform 4 entanglements, 4 phase shift, 4 Hadamard gates and 4 measurement in the computational basis. Thus, Bob has to perform 42,840 T gates, 106,488 two-qubit Clifford-group gates, 149,940 one-qubit Clifford-group gates, and 85,680 measurements. In our first fault-tolerant protocol, Alice needs a small quantum computer which consists of at least 22 qubits. She has to perform about 900,007.5 T gates, 20,485,885 two-qubit Clifford-group gates, 4,821,468.75 one-qubit Clifford-group gates, and 3,042,882.5 measurements in preparing encoded logical qubits. Bob has to perform about 899,640 T gates, 15,568,056 two-qubit Clifford-group gates, 3,534,300 one-qubit Clifford-group gates, and 2,441,880 measurements. In our BSA fault-tolerant protocol, Alice needs a buffer of at most 56 qubits and she does no quantum computation except receiving and sending qubits and running quantum error correction locally. Bob has to perform about 7,200,060 T gates, 163,887,080 two-qubit Clifford-group gates, 38,571,750 one-qubit Clifford-group gates, and 24,343,060 measurements in preparing encoded logical qubits. Bob also has to perform about 899,640 T gates, 15,568,056 two-qubit Clifford-group gates, 3,534,300 one-qubit Clifford-group gates, and 2,441,880 measurements for blind quantum computation.

Three tables show the comparison of resource consumption of performing a 10-bit QCLA adder using our fault-tolerant protocol. In the BFK basic protocol, compared to the target circuit, a 10-bit QCLA adder requires Bob more than 97 times as many quantum gates, and Alice has to prepare 85,715 qubits. Compared to the fault-tolerant quantum circuit of a 10-bit QCLA adder using 260 qubits, our first fault-tolerant protocol makes Alice perform more than 97 times as many quantum gates, but she needs only a small 22-qubit quantum computer. Compared to our first fault-tolerant protocol, our BSA protocol reduce the quantum computation overhead of Alice and requires Bob to execute 9 times more quantum gates.

%\begin{table}%
%\tbl{Resource consumption compared to the target circuit\label{tab:one}}{%
%\begin{tabular}{|l|r|r|r|r|}
%\hline
% & T gates& 2-qubit gates & 1-qubit gates & measurements \\\hline
%Decomposed circuit & 441 & 413 & 144 & \\\hline
%FT Decomposed circuit & 9,261 & 118,433 & 23,058 & 15,435 \\\hline
%BFK basic BQC & 42,840 & 106,488 & 149,940 & 85,680 \\\hline
%\end{tabular}}
%\end{table}%

\begin{table}%
\tbl{Resource consumption compared to the target circuit\label{tab:one}}{%
\begin{tabular}{|l|r|r|r|}
\hline
 & T gates& 2-qubit gates & 1-qubit gates \\\hline
FT circuit & 21x & 287x & 160x \\\hline
BFK basic BQC & 97x & 258x & 1041x \\\hline
Our 1st FT BQC (Bob) & 2,040x & 37,695x & 24,544x \\\hline
Our 1st FT BQC (Alice) & 2,041x & 49,603x & 33,482x \\\hline
Our BSA (Bob only) & 18,367x & 434,516x & 292,403x \\\hline
\end{tabular}}
\end{table}%

\begin{table}%
\tbl{Resource consumption compared to the fault-tolerant target circuit\label{tab:two}}{%
\begin{tabular}{|l|r|r|r|r|}
\hline
 & T gates& 2-qubit gates & 1-qubit gates & measurements \\\hline
Our 1st FT BQC (Bob) & 97x & 131x & 153x & 158x \\\hline
Our 1st FT BQC (Alice) & 97x & 173x & 209x & 197x \\\hline
Our BSA (Bob only) & 875x & 1,515x & 1,826x & 1,735x \\\hline
\end{tabular}}
\end{table}%

\begin{table}%
\tbl{Resource consumption compared to the BFK non-fault-tolerant BQC\label{tab:three}}{%
\begin{tabular}{|l|r|r|r|r|}
\hline
 & T gates& 2-qubit gates & 1-qubit gates & measurements \\\hline
Our 1st FT BQC (Bob) & 21x & 146x & 24x & 29x \\\hline
Our 1st FT BQC (Alice) & 21x & 192x & 32x & 36x \\\hline
Our BSA (Bob only) & 189x & 1,685x & 281x & 313x \\\hline
\end{tabular}}
\end{table}%

\section{Conclusions}
%Future work: implement with what model? FT-circuit or FT-mbqc(Topological)?
We propose two fault-tolerant blind quantum computation protocol which applies quantum error correction underneath the blind quantum computation. The first fault-tolerant protocol has two advantages. First, it provides more fault-tolerance. Second, the client prepares fewer qubits. However, the trade-off of the protocol is that the client needs a small quantum computer. Because of that, the client may leak some information which can be exploited by a side-channel attack. The second fault-tolerant protocol requires the client to do no quantum computation while preserving the two advantages of the first protocol. Thus, the side-channel attack that works against the first protocol does not work against this protocol. The second protocol has three trade-offs. First, the client has to buffer eight encoded logical qubits. Second, the server has to do more quantum computation. Third, this protocol uses more bandwidth of the quantum channel. As long as the server has enough quantum computation power and the quantum channel provides enough bandwidth, the client is required only to receive, buffer, and send qubits. We also provide a simple analysis on the resource consumption of fault-tolerant blind quantum computation. Compared to the BFK basic blind quantum computation, the overhead of our fault-tolerant blind quantum computation is increased by a constant factor.

%quantum teleportation
%buffer and shuffle
Our work can be further extended in the following ways. First, the idea of making the server prepare encoded logical qubits can be applied to some other blind quantum computation protocols, for example, the blind topological measurement-based quantum computation \cite{MF12}. Second, the optimization of quantum circuit for blind quantum computation is an important issue because the client's overhead is proportional to the size of the brickwork state. Third, quantum resources used in transmitting qubit via quantum teleportation should be considered because that is also a non-negligible overhead.

% Bibliography
\bibliographystyle{ACM-Reference-Format-Journals}
\bibliography{FTBQC-bibfile}
                             % Sample .bib file with references that match those in
                             % the 'Specifications Document (V1.5)' as well containing
                             % 'legacy' bibs and bibs with 'alternate codings'.
                             % Gerry Murray - March 2012

% History dates
%\received{February 2007}{March 2009}{June 2009}

\end{document}